\documentclass[a4paper,11pt]{article}
\usepackage{tocloft}
\pdfoutput=1
\usepackage{jcappub}
\usepackage{amssymb,amsmath,mathrsfs,enumerate}
\usepackage{graphicx,rotate,multicol}
\usepackage{float}
\usepackage{subcaption}

\usepackage{braket}

\usepackage{slashed}
\usepackage{mathtools}
\usepackage{multirow}

\allowdisplaybreaks

\title{\boldmath Neutrino Decoherence via Modified Dispersion}

\author{Bikash Kumar Acharya,}
\author{Indra Kumar Banerjee,}
\author{Ujjal Kumar Dey}
\affiliation{Department of Physical Sciences, Indian Institute of Science Education and Research Berhampur,\\Ganjam, Odisha, 760003, India}

\emailAdd{bikashka22@iiserbpr.ac.in}
\emailAdd{indrab@iiserbpr.ac.in}
\emailAdd{ujjal@iiserbpr.ac.in}

\abstract{We study in detail the effect of quantum decoherence in neutrino oscillations. We adopt a phenomenological approach that allows us to parametrize the energy dependence of the decoherence effects resulting from the modification of the neutrino dispersion relation. Using the open quantum system framework we derive decoherence parameters, which are usually connected to quantum gravitational effects. Furthermore, we study the sensitivity of decoherence on high-energy astrophysical neutrinos among all possible initial source compositions. We find that variation in the flux composition at neutrino telescopes can be a good probe to test such effects. Additionally, we show that a simple extension with heavy sterile neutrino decoherence produces verifiable signatures.}

\begin{document}
\maketitle
\flushbottom


\section{Introduction}
\label{sec:intro}
One of the long-standing challenges in fundamental physics is the reconciliation of quantum theory and gravitation. The behaviour of space-time at the tiniest scales remains one of the mysteries as to whether it is susceptible to fluctuations rooted in quantum gravity (QG) and other potential theories. Neutrino, an elementary particle in the Standard Model (SM) of particle physics, provides a great avenue to probe the structure of space-time. The discovery of neutrino oscillation about two decades ago was a watershed moment which established that neutrinos are massive and opened the window to a plethora of new investigations in elementary particle physics. The past two decades have witnessed tremendous experimental progress to develop a clearer picture of neutrino oscillation~\cite{Kajita:2016cak, McDonald:2016ixn} as well as the measurement of the associated parameters with great accuracy~\cite{deSalas:2020pgw, Esteban:2020cvm, Capozzi:2021fjo}. Current and forthcoming neutrino detectors and experiments are expected to shed light on a number of unanswered questions, such as exact neutrino mass and its ordering, the octant of the atmospheric mixing angle, and a possible CP violation in the neutrino sector, and to determine all oscillation parameters with extreme precision~\cite{SAGE:1999zrn, MiniBooNE:2012meu, SciBooNE:2011qyf, Borexino:2013zhu, DUNE:2020fgq, Hyper-Kamiokande:2018ofw, KamLAND:2014gul, SNO:2011hxd, MicroBooNE:2016pwy, T2K:2011qtm, IceCube:2013low, IceCube:2013cdw, ICAL:2015stm, RNO-G:2023kag, NOvA:2007rmc, Kamiokande:1986xwg, K2K:2000kji, ANTARES:1999fhm, MINOS:1998kez, Anderson:1998zza, CHOOZ:1997cow, Super-Kamiokande:1998wen, IceCube-Gen2:2020qha, DayaBay:2007fgu, Frejus:1994brq, RENO:2018dro, Andres:1999hm, DeBonis:2014jlo, KM3NeT:2018wnd, GRAND:2018iaj}.
From its source, after originating in a specific flavour, the neutrino propagates as a superposition of its mass eigenstates. During propagation, its mass eigenstates evolve with different frequencies, thus resulting in conversion from one flavour to another, known as neutrino oscillations~\cite{Super-Kamiokande:1998kpq, Berezinsky:2001uv}. Since according to SM neutrinos are stable and interact weakly, this quantum superposition remains intact over macroscopic length scales. The traditional study of neutrino oscillations assumes the neutrinos to remain isolated from its environment, and the oscillation effects are coherent. However, this coherence resulting from unhindered quantum superposition of neutrino eigenstates can be lost if neutrinos start interacting with the stochastic environment, thus resulting in dampening of oscillation probabilities, a phenomenon known as neutrino decoherence. Such effects in neutrino(s) can be observed by considering different physical origins~\cite{Hooper:2005jp,  Jones:2014sfa, Giunti:2003ax, Stuttard:2020qfv, DeRomeri:2023dht, Krueger:2023skk, Farzan:2008eg, Beuthe:2001rc, Ohlsson:2000mj, Kiers:1995zj}.
The coherence loss in neutrinos can originate predominantly in two ways, i.e. (i) the intrinsic decoherence which is a result of the broadening of the width of the wave packet, and (ii) the extrinsic decoherence, where the interaction of the neutrino subsystem with the environment is the main player. This work only investigates decoherence resulting from the latter type, adapting the open quantum system (OQS)~\cite{Breuer:2002pc} framework commonly used for decoherence studies. To describe this, we use the Gorini-Kossakowski-Sudarshan-Lindblad (GKSL) master equation~\cite{Gorini:1975nb, Lindblad:1975ef,  Chruciski:2017} which describes the OQS. Working in this framework and allowing the neutrino (sub)system to interact with the environment, we will study the decoherence effects on the oscillation phenomenon~\cite{Stuttard:2020qfv, Banerjee:2022slh, Gomes:2020muc, Buoninfante:2020iyr, Carpio:2018gum, Coloma:2018, Oliveira:2016asf, Fogli:2007tx, Morgan:2004vv, Lisi:2000zt, Benatti:2000ph, Gago:2000qc}. 
%

%
This work explores the possibility of neutrino and environment interaction in the vacuum due to quantum gravity effects, considering the fluctuating space-time as the stochastic background. Near the Planck scale, the fluctuations in the space-time metric would alter the conventional rules of quantum mechanics for the propagating neutrinos. As a result, uncertainties in neutrino energies and momentum will creep in. These uncertainties would modify the usual dispersion relation of the neutrino system, leading to one possible mechanism of coherence loss. We also compare the decoherence signatures with the current constraints and discuss the prospects of testing it in the forthcoming experiments. It is worth mentioning that neutrino propagation through matter effects also lead to significant changes in the oscillation probabilities. However, we mainly focus on the neutrinos originating from far away astrophysical sources and propagating astronomical distances through vacuum and hence do not consider matter effects. In this study, we focus on the new physics effects arising from neutrino and environment interactions at high energies, deliberately omitting the decoherence induced by wave-packet separation relevant at lower energies.
This article is organized as follows. In Sec.~\ref{sec:nuosc}, we briefly introduce neutrino oscillation to set our notations and conventions. In Sec.~\ref{sec:nuOQS}, we present the theoretical formalism of OQS and how neutrinos can be treated in this framework. We discuss in Sec.~\ref{sec:mdrmastereqn}, the phenomenological models and derive the modified master equation. Furthermore, we comment on energy dependence of decoherence parameters in Sec.~\ref{sec:energydep}. Afterwards, we discuss our results in Sec.~\ref{sec:resanddisc}. In addition, we briefly study $3+1$ flavour scenario in Sec.~\ref{sec:modelextend}. We finally summarise and conclude in Sec.~\ref{sec:sumandconc}.

\section{A quick look at Neutrino Oscillation}
\label{sec:nuosc}

This section briefly overviews the conventional treatment of neutrino oscillation to establish our notations and conventions. Neutrino flavour eigenstates can generally be expressed in terms of the neutrino mass eigenstates using the Pontecorvo-Maki-Nakagawa-Sakata (PMNS) mixing matrix,
\begin{equation}
    |\nu_\alpha\rangle = \sum_{a=1}^{n} U^*_{\alpha a} |\nu_a\rangle,
    \label{eq:f2m}
\end{equation}
where $n$ is the number of neutrino mass eigenstates, and $U$ is the PMNS matrix. For the standard three-neutrino case, $n = 3$ and $\alpha = e, \mu, \tau$. In the most general case, the unitary $n \times n$ PMNS matrix, which relates the flavour states $|\nu_\alpha\rangle$ ($\alpha = e, \mu, \tau, \ldots$) and the mass eigenstates $|\nu_a\rangle$ ($a = 1, 2, 3,\ldots, n$) is given by,
\begin{equation}
    U = \begin{pmatrix}
        U_{e1} & U_{e2} & U_{e3} & \ldots \\
        U_{\mu 1} & U_{\mu 2} & U_{\mu 3} & \ldots \\
        U_{\tau 1} & U_{\tau 2} & U_{\tau 3} & \ldots \\
        \vdots & \vdots & \vdots & \ddots
    \end{pmatrix}.
    \label{eq:umat}
\end{equation}
In this work, at first, we proceed in a three-flavour scenario followed by a $3+1$ scenario for neutrino oscillation and decoherence. For two-flavour studies, interested readers can see ~\cite{Ohlsson:2000mj, Giunti:2007ry}.
\subsection*{Three Flavour Neutrino Oscillation}
\label{subsec:3flav}
In the scenario of three-flavour neutrino oscillations, the unitary mixing matrix \(U\) takes the form of a \(3 \times 3\) unitary matrix. Three mixing angles and a complex phase characterize it and can be expressed as,
\begin{equation}
U =
\begin{pmatrix}
    U_{e1} & U_{e2} & U_{e3} \\
    U_{\mu1} & U_{\mu2} & U_{\mu3} \\
    U_{\tau1} & U_{\tau2} & U_{\tau3}
\end{pmatrix} =
\begin{pmatrix}
    c_{12}c_{13} & s_{12}c_{13} & s_{13}e^{-i\delta} \\
    -s_{12}c_{23} - c_{12}s_{23}s_{13}e^{i\delta} & c_{12}c_{23} - s_{12}s_{23}s_{13}e^{i\delta} & s_{23}c_{13} \\
    s_{12}s_{23} - c_{12}c_{23}s_{13}e^{i\delta} & -c_{12}s_{23} - s_{12}c_{23}s_{13}e^{i\delta} & c_{23}c_{13}
\end{pmatrix},
\label{eq:3fumat}
\end{equation}
where \(c_{ij} = \cos \theta_{ij}\), \(s_{ij} = \sin \theta_{ij}\), and \(\delta\) is the CP phase. The Schr\"{o}dinger equation represents the time evolution of the flavour states,
\begin{equation}
i \frac{\partial}{\partial t} 
\begin{pmatrix} 
| \nu_e(t) \rangle \\ 
| \nu_\mu(t) \rangle \\ 
| \nu_\tau(t) \rangle 
\end{pmatrix}
= H_f 
\begin{pmatrix} 
| \nu_e(t) \rangle \\ 
| \nu_\mu(t) \rangle \\ 
| \nu_\tau(t) \rangle 
\end{pmatrix}.
\label{eq:3fSE}
\end{equation}
The Hamiltonian in mass eigenbasis for the three-flavour scenario takes the form, 
\begin{equation}
H_m =
\begin{pmatrix}
    E_1 & 0 & 0\\
    0 & E_2 & 0\\
    0 & 0 & E_3
 \end{pmatrix} \rightarrow \frac{1}{2E}
\begin{pmatrix}
    0 & 0 & 0 \\
    0 & \Delta m^2_{21} & 0 \\
    0 & 0 & \Delta m^2_{31}
\end{pmatrix} \equiv 
\begin{pmatrix}
	0 & 0 & 0 \\
    0 & \Delta_{21} & 0 \\
    0 & 0 & \Delta_{31}
\end{pmatrix}
\label{eq:3fham}
\end{equation}
where $\Delta_{ij}=\Delta m^2_{ij}/2E$ and \(H_f = \tilde U H_m \tilde U^{-1}\) is the Hamiltonian matrix in flavour basis and \(H_m\) is the diagonal Hamiltonian that governs the time evolution of neutrino mass eigenstates with $\tilde U$ is unitary matrix. The energies of three different flavour neutrinos are assumed to be same here. In three-flavour neutrino oscillation paradigm, the probability of oscillation between different neutrino flavours, $\nu_\alpha$ and $\nu_\beta$, is given by,
\begin{align}
    P_{\nu_\alpha \rightarrow \nu_\beta} = \delta_{\alpha\beta} &- 4 \sum_{a,b=1}^{3} \operatorname{Re} (U^*_{\alpha a} U_{\beta a} U_{\alpha b} U^*_{\beta b}) \sin^2\left(\frac{\Delta m^2_{ab}L}{4E}\right) \nonumber \\
      &+ 2 \sum_{a,b=1}^{3} \operatorname{Im}(U^*_{\alpha a}U_{\beta a}U_{\alpha b}U^*_{\beta b}) \sin\left(\frac{\Delta m^2_{ab}L}{2E}\right).
      \label{eq:3fprob}
 \end{align}
Here, $\delta_{\alpha\beta}$ is the Kronecker delta, $L$ is the neutrino path length, and $E$ is the neutrino energy. The oscillation probability for the $\alpha = \beta$ case is known as survival probability, and $\alpha \neq \beta$ is termed as transition probability. Since in practice neutrinos are neither detected nor produced with sharp energy or with well defined propagation length, we have to average over the $L/E$ dependence and other uncertainties in the detection and emission. When neutrinos have length and energy spread $f(L,E)$, the oscillatory terms in Eq.~\eqref{eq:3fprob} fluctuate rapidly and can average out while they reach detectors on Earth. This highlights the need for Gaussian averaging, as the energy spread over long baselines disrupts phase coherence, causing flavor transitions to become probabilistic instead of oscillatory. The Gaussian averging is defined as,
\begin{equation}
\begin{array}{l@{\hspace{2cm}}l}
\displaystyle \langle P \rangle \equiv \int_{-\infty}^{\infty} P(x)\, f(x)\, dx, &
\displaystyle f(x) \equiv \frac{1}{\sigma \sqrt{2 \pi}} e^{-\frac{(x-\ell)^2}{2 \sigma^2}} 
\end{array}
\end{equation}
where $\ell \equiv \langle x \rangle$ and $\sigma \equiv \sqrt{\langle
(x - \langle x \rangle)^2 \rangle}$ are the expectation value and standard
deviation, respectively. By taking the Gaussian average of Eq.~\eqref{eq:3fprob} and using
$x \equiv \frac{L}{4E}$, we get,
\begin{equation}
\begin{split}
    \langle P_{\nu_\alpha \rightarrow \nu_\beta} \rangle = \delta_{\alpha\beta} &- 2 \sum_{a,b=1}^{3} \operatorname{Re}\left(U^*_{\alpha a}U_{\beta a}U_{\alpha b}U^*_{\beta b}\right) 
    \left(1 - \cos\left(\frac{\Delta m^2_{ab}L}{2E}\right) e^{-2\sigma^2(\Delta m^2_{ab})}\right) \\
    & + 2 \sum_{a,b=1}^{3} \operatorname{Im}\left(U^*_{\alpha a}U_{\beta a}U_{\alpha b}U^*_{\beta b}\right) \sin\left(\frac{\Delta m^2_{ab}L}{2E}\right) e^{-2\sigma^2(\Delta m^2_{ab})^2},
    \end{split}
    \label{eq:3flavprob}
\end{equation}
where $\sigma$ is the damping parameter responsible for damping the neutrino oscillation probabilities which can be parametrized by the decoherence parameter (described in later sections) following the equivalence established between the two as in Ref~\cite{Ohlsson:2000mj}\footnote{It is to be noted here that the left hand sides of both Eqs.~\eqref{eq:3flavprob} and \eqref{eq:3fdecohprob} denote the same quantity. However, in order to avoid confusion, we do not use $\langle...\rangle$ in the left hand side of Eq.~\eqref{eq:3fdecohprob}.}, which will damp the probability as,
\begin{equation}
\begin{split}
P_{\nu_\alpha \rightarrow \nu_\beta} = \delta_{\alpha\beta}& - 2 \sum_{a,b=1}^{3} \operatorname{Re}\left(U^*_{\alpha a}U_{\beta a}U_{\alpha b}U^*_{\beta b}\right) \left(1 - \cos\left(\frac{\Delta m^2_{ab}L}{2E}\right)e^{-\Gamma_{ab}(E)L}\right) \\& + 2 \sum_{a,b=1}^{3} \operatorname{Im}\left(U^*_{\alpha a}U_{\beta a}U_{\alpha b}U^*_{\beta b}\right) \sin\left(\frac{\Delta m^2_{ab}L}{2E}\right)e^{-\Gamma_{ab}(E)L}
\end{split}
    \label{eq:3fdecohprob}
\end{equation}
where $\Gamma_{ab}$ are the decoherence parameters which in turn will depend on the decoherence effects under consideration. For our numerical analysis we use the experimentally obtained best-fit values from the NuFit collaboration~\cite{Esteban:2024eli} given in Tab.~\ref{tab:table1}.
\begin{table}[ht]
\centering
\begin{tabular}{lcc}
\hline
\textbf{Parameter} & \textbf{Normal Ordering (NO)} & \textbf{Inverted Ordering (IO)} \\
\hline
$\sin^2 \theta_{12}$ & $0.308 \pm 0.012$ & $0.308 \pm 0.012$ \\
$\sin^2 \theta_{23}$ & $0.470^{+0.017}_{-0.013}$ & $0.550^{+0.012}_{-0.015}$ \\
$\sin^2 \theta_{13}$ & $0.02215 \pm 0.00057$ & $0.02231 \pm 0.00056$ \\
$\Delta m^2_{21}$ [$10^{-5}~\mathrm{eV}^2$] & $7.49 \pm 0.19$ & $7.49 \pm 0.19$ \\
$\Delta m^2_{3\ell}$ [$10^{-3}~\mathrm{eV}^2$] & $+2.513 \pm 0.021$ & $-2.484 \pm 0.020$ \\
$\delta_{\rm CP}$ [$^\circ$] & $212^{+26}_{-41}$ & $274^{+22}_{-25}$ \\
\hline
\end{tabular}
\caption{Neutrino oscillation parameters from NuFit-6.0 (2024), based on global fit with SK atmospheric data.}
\label{tab:table1}
\end{table}
Moreover, in this work we will consider just the central values of these parameters and the CP-violating phase taken to be zero.

\section{Neutrino as an Open Quantum System}
\label{sec:nuOQS}
In this section, we prepare the framework for the interaction of a neutrino subsystem with a given environment. As mentioned earlier, neutrino decoherence occurs when we consider a neutrino system coupled to an environment such as a reservoir or a bath. In this kind of scenario the Schr\"{o}dinger equation can not describe the evolution, as the coupling to the environment produces mixed quantum mechanical states even if the initial state is pure. Instead, we must use the GKSL master equation to describe system-environment interactions. The quantum states of neutrinos, pure and mixed, can be mathematically expressed using the density matrix, $\rho$, for a system of $j$ states of probability $p_j$ is given by: 
\begin{equation}
\label{eqn:denmatrix}
\rho(t) = \sum_j p_j \ket{\psi_j} \bra{\psi_j}.
\end{equation}
This matrix is Hermitian (\(\rho^\dagger = \rho\)), with positive eigenvalues and maintains a constant trace equal to one (\(\text{Tr} \, \rho = 1\)). The Liouville-von Neumann equation governs the dynamics of the neutrino density matrix:
\begin{equation}
\label{eq:liouveq1}
    \dot{\rho}(t) = -i [H_m, \rho],
\end{equation}
where $\rho$ and $H_m$ are the density matrix and the Hamiltonian of the neutrino subsystem, respectively. At this stage, we emphasize that decoherence occurs simultaneously with the oscillation of neutrinos as they propagate. Rather than halting the oscillation, decoherence modifies the oscillatory patterns. Given that neutrino oscillations are determined by the mass basis Hamiltonian, and since decoherence which is described by a dissipator that is expressed in terms of mass basis Hamiltonian, it follows that the phenomenon of decoherence must also be described by the mass basis Hamiltonian. To incorporate decoherence effects, an additional term \(D[\rho]\) is introduced in the Liouville equation \eqref{eq:liouveq1} for the neutrinos, which will allow transitions from pure to mixed quantum mechanical states, is given by,
\begin{equation}
\label{eq:liouveq2}
    \dot{\rho}(t) = -i [H_m, \rho] - D[\rho],
\end{equation}
where $D$ is a superoperator that contains all the information necessary to characterize the interaction of the neutrino subsystem with the environment. This can be described as ~\cite{Mavromatos:2004sz, BalieiroGomes:2018gtd},
\begin{equation}
D[\rho] = \frac{1}{2} \sum_{k=0}^{N^2 - 1} \left( [V_k, \rho V_k^\dagger] + [V_k \rho, V_k^\dagger] \right),
\label{eq:dissipator}
\end{equation}
where \(V_k = \sum_{i=0}^8 v_{ki}\lambda_i\) (with $k = 0, 1, \ldots, N^2 - 1$  where $N$ is the dimension of Hilbert space and in case of three flavour neutrinos $N=3$) represent the Lindblad operators, which are defined to satisfy the orthonormality and tracelessness conditions, \(\mathrm{Tr}(V_k) = 0\) and \(\mathrm{Tr}(V_k^\dagger V_j) = \delta_{kj}\). In general, these operators may exhibit explicit time dependence, \(V_k \rightarrow V_k(t)\), reflecting the temporal variation of the system-environment interaction. For two and three-flavour neutrino mixing cases, $V_k$s are represented by the Pauli and Gell-Mann matrices, respectively. Assuming \( V_k \)'s to be Hermitian ($V_k^\dagger = V_k$) and also commuting with the Hamiltonian ($[H_m, V_k] = 0$), the dissipator takes the form:
\begin{equation}
    D[\rho] = \sum_{k=1}^{} [V_k, [V_k, \rho]] = \sum_{k}^{} (\rho V_k^2 + V_k^2 \rho - V_k\rho V_k),
\end{equation}
As we have mandated all the operators \( \rho \), \( H_m \), \( D_a \) to be Hermitian, they can be spanned in terms of Pauli and Gell-Mann matrices depending upon the level of the system. Here, we consider a simple transformation \( H_m \rightarrow H'_m = H_m - \frac{1}{3} \text{Tr} (H_m) I_3 \) where \( I_3 \) is the \(3 \times 3\) identity matrix. Since \( H_m \) and \( H'_m \) differ by a trace, the resulting global phase is irrelevant as it will not affect the probabilities. From now onwards, we use $H$ instead on $H_m$ or $H_m^{\prime}$.

\subsection*{Three-flavour Decoherence}
\label{subsec:3fdecohform}
To study the decoherence in neutrinos with three flavours, we follow the methodology outlined in Ref.~\cite{Ohlsson:2000mj} and span the relevant observables with Gell-Mann matrices as follows,
\begin{align}
\rho   &= \frac{1}{3} \rho_0\lambda_0 + \frac{1}{2} \sum_{k=1}^{8} \rho_k\lambda_k,  \label{eq:density} \\
H      &= \mathbf{\textit{h}}_\nu \lambda_\nu, \quad \nu = 0,1,\ldots,8,             \label{eq:hamiltonian} \\
V_k    &= \sum_{i=0}^8 v_{ki}\lambda_i, \quad k = 0, 1, \ldots, 8,                  \label{eq:potential}
\end{align}
where \(\lambda_{k}\) represents the Gell-Mann matrices that satisfies \(\lambda_k^\dagger = \lambda_k\), \([\lambda_a, \lambda_b] = 2if_{abc}\lambda_c\), with structure constants \(f_{abc} = -\frac{i}{4} \text{Tr} (\lambda_a [\lambda_b, \lambda_c])\) with \(\lambda_0\) being the \(3 \times 3\) identity matrix. To have a complete unitary description of neutrino dynamics, it is essential to satisfy the necessary conditions for a density matrix, including trace preservation and complete positivity. The identity matrix ($\lambda_0$) plays a critical role in this framework. Additionally, to effectively describe the three-neutrino quantum system, we require a complete nine-element Hermitian basis $\{\lambda_0,\lambda_1,\dots,\lambda_8\}$ to expand each observable. The inclusion of $\lambda_0$ is crucial as it ensures trace preservation ($\operatorname{Tr}\rho = 1$) and allows for the representation of any $3 \times 3$ Hermitian operator, such as the density matrix $\rho$, Hamiltonian $H$, or interaction potential $V_k$, as a linear combination of these basis matrices as outlined in Eq.~\eqref{eq:density}, Eq.~\eqref{eq:hamiltonian} and Eq.~\eqref{eq:potential} respectively. Notably, since $\lambda_0$ commutes with all $\lambda_i$, it does not interfere with the existing commutation relations, maintaining the necessary structure for neutrino dynamics analysis. Substituting above equations into the GKSL master Eq.~\eqref{eq:liouveq2} we obtain:
\begin{equation}
\dot{\rho}_{\mu}(t) = f_{{\mu}ij} H_i \rho_j(t) + D_{\mu\nu} \rho_\nu(t).
\end{equation} 
This is a system of nine coupled differential equations. The form of Hamiltonian is given as $ H \equiv (H_0, 0, 0, H_3, 0, 0, 0, 0, H_8)$ with $H_0 = \Delta_{21} + \Delta_{31}$, $H_3 = -\Delta_{21}$, $H_8 = (\Delta_{21} - 2\Delta_{31})/\sqrt{3}$.
The dissipator in three-flavour case takes the form,
\begin{align}
    D_{\mu\nu} = \begin{pmatrix}
    -\Gamma_{00} & \beta_{01} & \beta_{02} & \beta_{03} & \beta_{04} & \beta_{05} & \beta_{06} & \beta_{07} & \beta_{08} \\
    \beta_{01} & -\Gamma_{11} & \beta_{12} & \beta_{13} & \beta_{14} & \beta_{15} & \beta_{16} & \beta_{17} & \beta_{18} \\
    \beta_{02} & \beta_{12} & -\Gamma_{22} & \beta_{23} & \beta_{24} & \beta_{25} & \beta_{26} & \beta_{27} & \beta_{28} \\
    \beta_{03} & \beta_{13} & \beta_{23} & -\Gamma_{33} & \beta_{34} & \beta_{35} & \beta_{36} & \beta_{37} & \beta_{38} \\
    \beta_{04} & \beta_{14} & \beta_{24} & \beta_{34} & -\Gamma_{44} & \beta_{45} & \beta_{46} & \beta_{47} & \beta_{48} \\
    \beta_{05} & \beta_{15} & \beta_{25} & \beta_{35} & \beta_{45} & -\Gamma_{55} & \beta_{56} & \beta_{57} & \beta_{58} \\
    \beta_{06} & \beta_{16} & \beta_{26} & \beta_{36} & \beta_{46} & \beta_{56} & -\Gamma_{66} & \beta_{67} & \beta_{68} \\
    \beta_{07} & \beta_{17} & \beta_{27} & \beta_{37} & \beta_{47} & \beta_{57} & \beta_{67} & -\Gamma_{77} & \beta_{78} \\
    \beta_{08} & \beta_{18} & \beta_{28} & \beta_{38} & \beta_{48} & \beta_{58} & \beta_{68} & \beta_{78} & -\Gamma_{88} 
    \end{pmatrix}.
    \label{eq:dissipator}
\end{align}
Now, we impose certain physical constraints to reduce the number of free parameters in the dissipator, which include probability conservation, which implies $D_{{\mu}0} = D_{0{\nu}} = 0$; trace-preservation, i.e., to ensure that all entries are real entries  and the $\Gamma_i$s are positive so that $\text{Tr}(\rho(t)) = 1$, and since $f_{0ij} = 0$ and $D_{0\nu} = 0$, we obtain $\dot{\rho}_0(t) = 0 \Rightarrow \rho_0(t) = 1$.
In theory, the possibility of energy non-conservation within the neutrino subsystem exists, although the energy is inherently conserved for the entire system. However, we strictly adhere to the energy conservation condition within the neutrino subsystem for our analysis. Apart from this, conditions like entropy increase and the complete positivity of the neutrino density matrix further simplify the dissipator matrix. For simplicity, in this work, we consider a diagonal dissipator of the form as,
\begin{equation}
\label{eq:diagonaldissipator}
D_{\mu\nu} = -\text{diag}(0, \Gamma_{21}, \Gamma_{21}, 0, \Gamma_{31}, \Gamma_{31}, \Gamma_{32}, \Gamma_{32}, 0),
\end{equation}
where the elements of the first row and column vanish, and new variables are introduced, specifically $\Gamma_{11} = \Gamma_{22} \equiv \Gamma_{21}$, $\Gamma_{44} = \Gamma_{55} \equiv \Gamma_{31}$, and $\Gamma_{66} = \Gamma_{77} \equiv \Gamma_{32}$. These parameters will depend on the decoherence mechanism under consideration.
%
\section{Modified Dispersion Relation: Decoherence and Master Equation}
\label{sec:mdrmastereqn}
%
Many theories on quantum gravity predict a minimum measurable length, which in a sense `pixelate' the space-time. As a result, space and time, which are continuous, realize a quantum nature. There are many consequences of this effective `quantum' nature of space-time, such as interactions of propagating particles with the vacuum, fluctuations of the metric, etc. As a result, the energy and momentum of a system in vacuum acquire some uncertainty, which in turn modifies the dispersion relation. In the subsequent part of this section, we explain the modified dispersion relation (MDR) arising from different quantum gravity theories and how they induce decoherence for neutrinos propagating in this `foamy' space-time.

\subsection{MDR-Induced Decoherence}
\label{subsec:moddisprel}
The fluctuation of the metric due to the foamy nature of the space-time can be expressed as~\cite{Ng:1993jb},
\begin{equation}
\delta g_{\mu\nu}\gtrsim\left(\dfrac{l_P}{l}\right)^a,
\end{equation}
where $l_P$ is the Planck length and $a$ depends on the particular theory that leads to the fluctuation~\cite{Ng:1993jb,Ng:2000fq,Ng:1999hm,Susskind:1994vu,Gambini:2007vn,Ng:2003ag}. This kind of metric perturbation can arise from the uncertainty in length measurement of the form $\delta l \gtrsim l^{1-a}l_P^a$ where the space-time fluctuation and the uncertainty in length are related by $\delta g = \delta l^2/l^2$. Furthermore, the uncertainty in the space-time can in turn be translated into the uncertainty in the energy-momentum tensor where the two can be related by $T^{\mu\nu}\delta g_{\mu\nu} = g_{\mu\nu}\delta T^{\mu\nu}.$ Using this relation, the uncertainty on the momentum and energy can be expressed as,
\begin{align}
\delta E&\gtrsim E\left(\dfrac{E}{E_P}\right)^a,\\
\delta p&\gtrsim p\left(\dfrac{p}{m_P c}\right)^a,
\label{eq:modep}
\end{align} 
where $m_P$ is the Plank mass, $E_P$ is the Plank energy, and $c$ is the speed of light in vacuum. Due to these uncertainties in the momentum and the energy of a system, the dispersion relation, i.e., the relation between a particle's mass, energy, and momentum, gets modified and takes the form,
\begin{align}
(E+\delta E)^2=(p+\delta p)^2+m^2.
\label{eq:mdr1}
\end{align}
Combining Eqs.~\eqref{eq:modep} and~\eqref{eq:mdr1}, we get a simplified energy equation for a particle, i.e.,
\begin{align}
E\approx p+\dfrac{m^2}{2p}-m^2\left(\dfrac{E^{a-1}}{E^a_P}\right).
\label{eq:mdr2}
\end{align}
It is worth mentioning here that in obtaining the above relation, we have ignored terms $\mathcal{O}((E/E_P)^2)$ and higher. Though we are working in an ultra-relativistic regime, the energy of the particle in realistic cases is always many orders of magnitude smaller than the Planck energy. It is also to be noted here that for an ultra-relativistic free particle, the energy-momentum relation takes the form,
\begin{align}
E\approx p+\dfrac{m^2}{2p}.
\label{eq:dr}
\end{align}
Therefore, we can already compare Eqs.~\eqref{eq:dr} and~\eqref{eq:mdr2} to see that the energy and momentum uncertainties due to the fluctuating space-time create additional terms and lead to a modified dispersion relation. This modification may lead to effects such as decoherence which will be discussed subsequently.
In absence of any modified dispersion relation ($a\rightarrow\infty$), the Hamiltonian for three neutrino system takes the form given in Eq.~\eqref{eq:3fham}. However, in case the dispersion relation is of the form of Eq.~\eqref{eq:mdr2}
the Hamiltonian gets modified and takes the form,
\begin{align}
H=H_0-\dfrac{E^{a-1}}{E_P^a}\begin{pmatrix}
0 & 0 & 0\\
0 & \Delta m_{21}^2 & 0\\
0 & 0 & \Delta m_{31}^2
\end{pmatrix}.
\label{eq:totalham}
\end{align} 
Due to the $E_P$ suppression in the second term we can think of this piece as a perturbation or `interaction'-type Hamiltonian. This is also motivated by the fact that since the background space-time is now pixelated, it interacts with the neutrinos similarly to an environmental interaction. It is to be noted that this interaction Hamiltonian can be expressed as,
\begin{align}
H_\text{I}= - 2E \dfrac{E^{a-1}}{E_P^a}H_0.
\label{eq:interactionham}
\end{align}
In the next section, we will derive the modified version of the master equation that incorporates the new dispersion relation. 

\subsection{MDR-Induced Master Equation}
%
In this section, we derive the Lindblad-type master equation in the context of MDR following the prescription of~\cite{Petruzziello:2020wkd}. We start from the canonical Schr\"odinger equation without any loss of generality, given by
\begin{equation}
 i \hbar \partial_t |\psi\rangle  = H|\psi\rangle.
\label{schroeqn}
\end{equation}
Due to MDR one can achieve an effective Hamiltonian of the form,
\begin{equation}
\label{eq:Heff}
H_{\mathrm{eff}} = H_{\mathrm{0}} + H_\text{I}.
\end{equation}
where $H_0$ is the Hamiltonian in the standard case, and $H_\text{I}$ arises due to the interaction between the system and the environment. In this case, since the MDR is motivated by quantum gravity effects, we can assume that the environment is the vacuum itself which is interacting with the system through quantum gravitational effects. 
The time evolution of neutrino(s) as an open quantum system experiencing decoherence is given by Eq.~\eqref{eq:liouveq1}. Since the neutrino Hamiltonian is now being modified because of coupling to its environment given by Eq.~\eqref{eq:Heff}, Eq.~\eqref{eq:liouveq1} now takes the form,
\begin{equation}
\label{eq:effgksl}
\dot{\rho}(t) = -i \left[ H_{\mathrm{eff}}, \rho(t) \right] = -i \left[ H_{\mathrm{0}} + H_{\mathrm{I}}, \rho(t) \right].
\end{equation}
It is to be noted here that the above equation denotes the unitary evolution of the entire system, i.e., the neutrino and the environment. However, in this study, we are interested in finding the decoherence of the neutrino sub-system that originates from the effective non-unitarity induced by the MDR.
Therefore, to solve this equation for the density matrix neutrino subsystem, we move to the interaction picture, where the eigenstate which was previously in the Schr\"odinger picture, now takes the form,
\begin{equation}
\label{eq:intpsi}
|\psi\rangle = e^{-i H_{\mathrm{0}} t} \, |\psi_{\mathrm{I}}\rangle.
\end{equation}
Following the interaction picture prescription, the Schr\"odinger equation reduces to,
\begin{equation}
\label{eq:transham}
i \frac{\partial}{\partial t} |\psi_{\mathrm{I}}\rangle = H'_{\mathrm{I}} |\psi_{\mathrm{I}}\rangle, \quad \text{where} \quad H'_{\mathrm{I}} = e^{i H_{\mathrm{0}} t} H_{\mathrm{I}} e^{-i H_{\mathrm{0}} t}
\end{equation}
and hence, the Liouville-von Neumann equation i.e., Eq.~\eqref{eq:liouveq1}, becomes,
\begin{equation}
\label{eq:intgksl}
\dot{\rho}'(t) = -i \left[ H'_{\mathrm{I}}(t), \rho'(t) \right].
\end{equation}
Note that from here onward the prime represents the respective quantities in the interaction picture. However, since $[H_\mathrm{I},H_0]=0$, one can understand that $H_\mathrm{I} = H'_\mathrm{I}$.
A precise solution to this equation can be obtained by integrating both sides over the interval $[0,t]$, yielding,
\begin{equation}
\label{eq:intgkslintegrand}
\int_{0}^{t} \dot{\rho}'(t') \, dt' = -i \int_{0}^{t} \left[ H_{\mathrm{I}}, \rho'(t') \right] dt',
\end{equation}
which results in,
\begin{equation}
\label{eq:intgkslintegrated}
\rho'(t) = \rho'(0) - i \int_{0}^{t} \left[ H_{\mathrm{I}}, \rho'(t') \right] dt'.
\end{equation}
Now, substituting Eq.~\eqref{eq:intgkslintegrated} into the r.h.s. of Eq.~\eqref{eq:intgksl}, we get,
\begin{equation}
\label{eq:modifiedgksl}
\dot{\rho}'(t) = -i \left[ H_{\mathrm{I}}, \rho'(0) \right] 
- \int_{0}^{t} \left[ H_{\mathrm{I}}, \left[ H_{\mathrm{I}}, \rho'(t') \right] \right] dt'.
\end{equation}
Here, we use the Born-Markov approximation\footnote{Since, the effects induced by MDR are weak, hence, Born-Markov approximation can be used which governs for weak system-environment interaction.}~\cite{Petruzziello:2020wkd, Xu:2020pzr, Kolovsky:2020} to let the density matrix of the r.h.s. of the above equation depend on $t$ rather than $t'$.
\begin{equation}
\label{eq:bmapprox}
\dot{\rho}'(t) = -i \left[ H_{\mathrm{I}}, \rho'(0) \right] 
- \int_{0}^{t} \left[ H_{\mathrm{I}}, \left[ H_{\mathrm{I}}, \rho'(t) \right] \right] dt'.
\end{equation}
At this stage, we note that Quantum Gravity (QG) theories must include space-time fluctuations at the Planck scale. This can be done by introducing a fluctuating deformation parameter to address the fluctuating minimum length scale, requiring a stochastic treatment to capture the complexities involved. At this point, while averaging the fluctuations is a viable option, we have chosen not to pursue that approach in this study.
Since the first term on the r.h.s. of Eq. \eqref{eq:bmapprox} is commutator consists of time independent entities, it generates a global phase term in the neutrino oscillation probabilities and can therefore be neglected. Substituting $H_{\mathrm{I}}$ in terms of $H_{\mathrm{0}}$ from Eq.~\eqref{eq:interactionham} in Eq.~\eqref{eq:bmapprox} in the second term, we get,
\begin{equation}
\label{eq:gkslfinalmodified}
\dot{\rho}'(t) = -\kappa \left[ H_0, \left[ H_0, \rho'(t) \right] \right],
\end{equation}
where $\kappa = 4 E^2 t \left(\dfrac{E^{a-1}}{E_P^a}\right)^2 $.
Now, we can move back to  Schr\"odinger representation as follows,
\begin{equation}
\label{eq:den_matrix_Schropic}
\rho(t) = e^{-\frac{i H_0 t}{\hbar}} \rho'(t) e^{\frac{i H_0 t}{\hbar}},
\end{equation}
which allows us to write the Liouville-von Neumann equation  given by Eq. \eqref{eq:liouveq1}  now turned to GKSL master equation with the desired dissipator term, and we get:
\begin{equation}
\label{eq:newlindblad}
\dot{\rho}(t) = -i \left[ H_0, \rho(t) \right] - 4 E^2 t \left( \dfrac{E^{a-1}}{E_P^a} \right)^2 \left[ H_0, \left[ H_0, \rho(t) \right] \right].
\end{equation}
This is the new master equation, which incorporates the 
decoherence effects arising because of the modified dispersion relation. It should be noted here the $\rho$ in the above above expression is the density matrix of the neutrino sub-system with the Born-Markov approximation in the Schr\"odinger representation and hence should not be confused with the density matrix of the total closed system expressed in Eq.~\eqref{eq:effgksl}. From Eq.~\eqref{eq:newlindblad}, we can get nine coupled differential equations for neutrino density matrix as given by,
\begin{subequations}
\label{coupledeqs}
\begin{align}
\dot{\rho}_0(t) &= 0\,, \label{eq:coupledeqs:d11} \\[3mm]
\dot{\rho}_1(t) &= \frac{\Delta m^2_{21}}{2E} \rho_2(t) - \frac{(\Delta m^2_{21})^2 t}{E^{2-2a}E_P^{2a}} \rho_1(t)\,,  \\[3mm]
\dot{\rho}_2(t) &= -\frac{\Delta m^2_{21}}{2E} \rho_1(t) - \frac{(\Delta m^2_{21})^2 t}{E^{2-2a}E_P^{2a}} \rho_2(t)\,,  \\[3mm]
\dot{\rho}_3(t) &=  0\,, \label{eq:coupledeqs:d33} \\[3mm]
\dot{\rho}_4(t) &= \frac{\Delta m^2_{31}}{2E} \rho_5(t) - \frac{(\Delta m^2_{31})^2 t}{E^{2-2a}E_P^{2a}} \rho_4(t)\,,  \\[3mm]
\dot{\rho}_5(t) &= -\frac{\Delta m^2_{31}}{2E} \rho_4(t) - \frac{(\Delta m^2_{31})^2 t}{E^{2-2a}E_P^{2a}} \rho_5(t)\,,  \\[3mm]
\dot{\rho}_6(t) &= \frac{\Delta m^2_{32}}{2E} \rho_7(t) - \frac{(\Delta m^2_{32})^2 t}{E^{2-2a}E_P^{2a}} \rho_6(t)\,,  \\[3mm]
\dot{\rho}_7(t) &= -\frac{\Delta m^2_{32}}{2E} \rho_6(t) - \frac{(\Delta m^2_{32})^2 t}{E^{2-2a}E_P^{2a}} \rho_7(t)\,,  \\[3mm]
\dot{\rho}_8(t) &=  0\,. \label{eq:coupledeqs:d88}
\end{align}
\end{subequations}
The r.h.s. of Eqs.~\eqref{eq:coupledeqs:d11}, \eqref{eq:coupledeqs:d33}, and \eqref{eq:coupledeqs:d88} vanishes due to the specific form of the disspator chosen in Eq.~\eqref{eq:diagonaldissipator}. Furthermore, it can be seen from Eq.~\eqref{eq:newlindblad} that it takes different forms depending on the value of $a$, which is an input from the specific theory of quantum gravity that causes the modification to the dispersion relation. In this study,  we remain agnostic about any such theories of quantum gravity. Later we are going to consider a few specific values of $a$ and compare their effects. At this point, we note that,  the first terms on r.h.s. of Eq.~\eqref{eq:newlindblad} and Eqs.~\eqref{coupledeqs} account for trivial unitary dynamics of neutrino oscillations while the second terms govern neutrino-environment interactions due to MDR\footnote{In the second term of the r.h.s of Eq.~\eqref{eq:newlindblad}, the variable $t$ will denote the propagation time of the neutrinos. Since in this work we are mostly dealing with ultra-relativistic neutrinos, the propagation time $t$ can be replaced with the propagation length $L$ as shown in Eq.~\eqref{eq:decohpara}.}. Moreover, the second terms on the r.h.s. of Eq.~\eqref{eq:newlindblad} and Eqs.~\eqref{coupledeqs} give insights about the form of decoherence strength parameter. Its explicit time-dependence comes through the dependence on $\rho$ itself which is further discussed in Sec.~\ref{subsec:decohparae}. 
\section{Energy Dependence of Decoherence}
\label{sec:energydep}

We now consider some phenomenological models crucial to decoherence studies. The general open quantum system approach outlined in Sec.~\ref{sec:nuOQS} does not implicitly consider the energy dependence producing the decoherence effects. Still, it can be introduced through the free parameters present in the dissipator. Currently, no generally accepted QG theory exists to shed light on energy dependence. Hence, we follow the phenomenological route to introduce a general way of how the decoherence strength $\Gamma_{ij}$ varies with energy. This work assumes a power law dependence on energy, in the spirit of Refs.~\cite{Stuttard:2020qfv,  Ternes:2025mys, Banerjee:2022slh, DeRomeri:2023dht},
\begin{equation}
\Gamma_{ij}(E) = \Gamma_{ij}(E_0) \left( \frac{E}{E_0} \right)^n \, ,
\label{eq:powerlaw}
\end{equation}
where $E_0$ is a constant energy scale which we set to $E_0 = 1$~GeV, and $n$ is a power-law index to be tested experimentally. While $\Gamma_{ij}$ appears time-independent in Eq.~\eqref{eq:powerlaw}, it is important to note that an explicit time dependence may emerge if the environment possesses stochastic or dynamical properties, as determined by the specific model under consideration. The parameter $n$ can take both positive and negative values, including zero. We discuss a few implications depending on this as follows.
\begin{enumerate}
\item{{\bf Lightcone Fluctuating Model:}} ($n = -2$) \\
The case directly follows from the gravitationally induced decoherence that suggests a specific coupling between the neutrinos and the environment inspired by general relativity and linearised gravity
respectively ~\cite{Domi:2024ypm}. Also, such dependence may arise from space-time fluctuations assuming gravity is a quantum force ~\cite{Stuttard:2021uyw}. The reported bounds on decoherence parameters obtained by the former study go as low as $\sim 10^{-25}$ GeV for the vacuum oscillation case.
\item{{\bf Lorentz Invariant Model:}} ($n= -1$) \\
This dependence decisively mirrors the oscillation energy dependence, and it has been definitively established that Lorentz invariance is preserved when $\Gamma_{ij}$ is proportional to $1/E$ ~\cite{Lisi:2000zt}. It can be noted that such dependence is phenomenologically similar to that of invisible neutrino decay~\cite{Stuttard:2020qfv}. Importantly, Super-Kamiokande  bounds on decoherence parameters for this model is approximately $\sim 10^{-21}$ GeV~\cite{Super-Kamiokande:1998wen}.
\item{{\bf Energy Independent Model:}} ($n = 0$)\\
If the strength of decoherence $\Gamma_{ij}$ is considered to be independent of energy, we observe that the oscillation probabilities given in Eq.~\eqref{eq:3flavprob} equilibrate at a value of 1/3 for distances greater than the coherence length defined in Sec.~\ref{subsec:cohlen}. Prospects of this model have been studied in the context of forthcoming reactor experiments such as DUNE and ESSnuSB. Current upper bounds on this parameter from these two experiments stand approximately at $\sim 10^{-23}$ GeV and $\sim 10^{-24}$ GeV respectively~\cite{BalieiroGomes:2018gtd, ESSnuSB:2024yji}.
\item{{\bf Cross-Section Model:}} ($n = 1$) \\
Such scaling in energy lies in line with the cross-section dependence. In Ref.~\cite{Liu:1997km}, such dependence was considered in the backdrop of solving the solar neutrino puzzle. The bounds obtained on $\Gamma_{ij}$'s are $\sim 10^{-10}$ GeV.
\item{{\bf String Motivated Model:}} ($n = 2$) \\
It has been suggested that $\Gamma_{ij}$ may scale with $E^2$,
particularly within the context of string theory
\cite{Ellis:1992eh, Ellis:1993wh, Gambini:2003pv, Ellis:1995xd, Ellis:1996bz, Ellis:1997jw, Ellis:2000dy}. Following this model, IceCube has set upper bounds, and they are $\sim 10^{-27}$ GeV
\cite{ICECUBE:2023gdv, Coloma:2018}.
\end{enumerate}

\section{Results and Discussion}
\label{sec:resanddisc}
In this section, we discuss relevant observables and parameters pertaining to this study and the effects of decoherence arising from MDR given in Sec.~\ref {sec:mdrmastereqn}.  In addition to the standard three-flavour case, we also consider adding one sterile neutrino and study the impact of this extension guided by decoherence.

\subsection{Observables and Relevant Parameters}
\label{subsec:obsandpara}

%
The list of observables emanating from this model, which is of interest to study, as well as the relevant parameters that affect them, is as follows.
\begin{enumerate}
\item \textbf{Decoherence Strength or Relaxation:} Three independent strength may arise from this model, i.e., $\Gamma_{21}(E)$, $\Gamma_{31}(E)$ and $\Gamma_{32}(E)$. As a result, different modifications of oscillation patterns will occur after different length of propagation which can in principle be probed by forthcoming experiments.
\item \textbf{Oscillation Probabilities:} Studying the neutrino survival and transition probabilities can provide insights on decoherence strength parameters, mass hierarchies and help to testify to the flavour equilibration notion and quantify distortions on oscillations creeping in due to decoherence.
\item \textbf{Flavour Composition:} One of the outstanding observables to probe quantum decoherence is the flavour composition at neutrino telescopes described more lucidly in Sec. ~\ref{subsec:flavcomp}. This can tell us about the source and the effect of decoherence with respect to a vast range of energy and the path traversed.
\end{enumerate}
%
%
\subsection{Decoherence Strength Parameters}
\label{subsec:decohparae}
%
The effect of decoherence on oscillating neutrinos is quantified by the decoherence strength parameters $\Gamma_{ij}(E)$. The form of the decoherence strength parameters can be evaluated from Eq.~\eqref{eq:newlindblad} and can be expressed as,
\begin{align}
\Gamma_{ij}(E) &= \dfrac{(\Delta m^2_{ij})^2 L}{E^{2-2a}E_P^{2a}}.
\label{eq:decohpara}
\end{align}
It is evident from the previous expression, different values of $a$ correspond to different energy dependence ($n=2a-2$) which has been briefly explained in Sec.~\ref{sec:energydep}. Also, it is to be noted that, once the neutrino source is fixed, the dissipator effectively becomes time (or length)-independent.
\begin{figure}[t]
	\centering
	\includegraphics[scale=1.1]{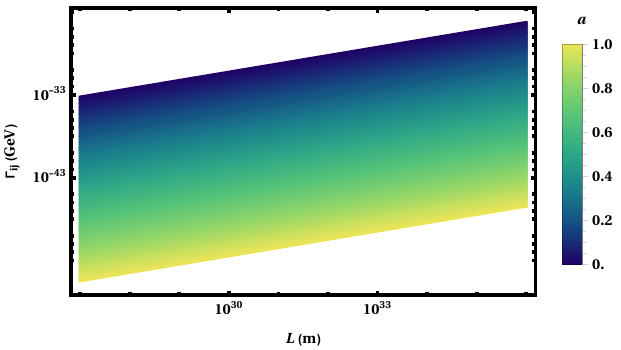} 
	\caption{The dependence of the decoherence strength parameter on propagation length with varying $a$ is shown.} 
	\label{fig:paraplot}
\end{figure}
In the three-flavour paradigm, we will have three distinct decoherence parameters. From Fig.~\ref{fig:paraplot}, it can be understood that although the behaviour of the decoherence parameters depends on the value of $a$, for $a>0$, the value of decoherence parameters will be suppressed by $E_P$, which is expected as this effect arises from the aspect of quantum gravity. For $a\rightarrow 1$, the $\Gamma_{ij}$ is suppressed for all energy scales. This is because the fluctuation in the metric is most suppressed at $a=1$ and most dominant at $a=0$. Therefore, for the impact of the decoherence to be visible, the neutrino propagation length must be very large to accumulate these effects. Furthermore, for $a>0$, the value of the decoherence parameters increases with decreasing energy of the neutrinos. Hence, far-away sources with detectable yet relatively low-energy neutrinos are our best bet to observe these effects. The current upper bounds on this parameter from various ongoing experiments have been discussed in Sec.~\ref{sec:energydep}.
%
\subsection{Coherence Length}
\label{subsec:cohlen}
In neutrino decoherence studies, the coherence length $l_{\mathrm{coh}}$ is a measure of the propagation length after which the survival and transition probability amplitudes of the neutrinos are suppressed by a factor of $e$ (see  Eq.~\eqref{eq:3fdecohprob} for a quick reference) and can be expressed as the inverse of the decoherence strength, i.e., $l_{\text{coh}} = 1/ \Gamma(E)$.
%
%
Since, for all values of $a$, except $a=1$, we have negative energy dependency of the decoherence parameter, and hence the coherence length increases with energy, as is evident from Eq.~\eqref{eq:decohpara}. This suggests that the effects of quantum gravity are reasonably prominent over a broad spectrum of energy ranging from a few fractions of meV to PeV. This helps us to pick the right kind of sources to look for while studying decoherence.
\begin{figure}[htbp]
\centering
\includegraphics[scale=0.26]{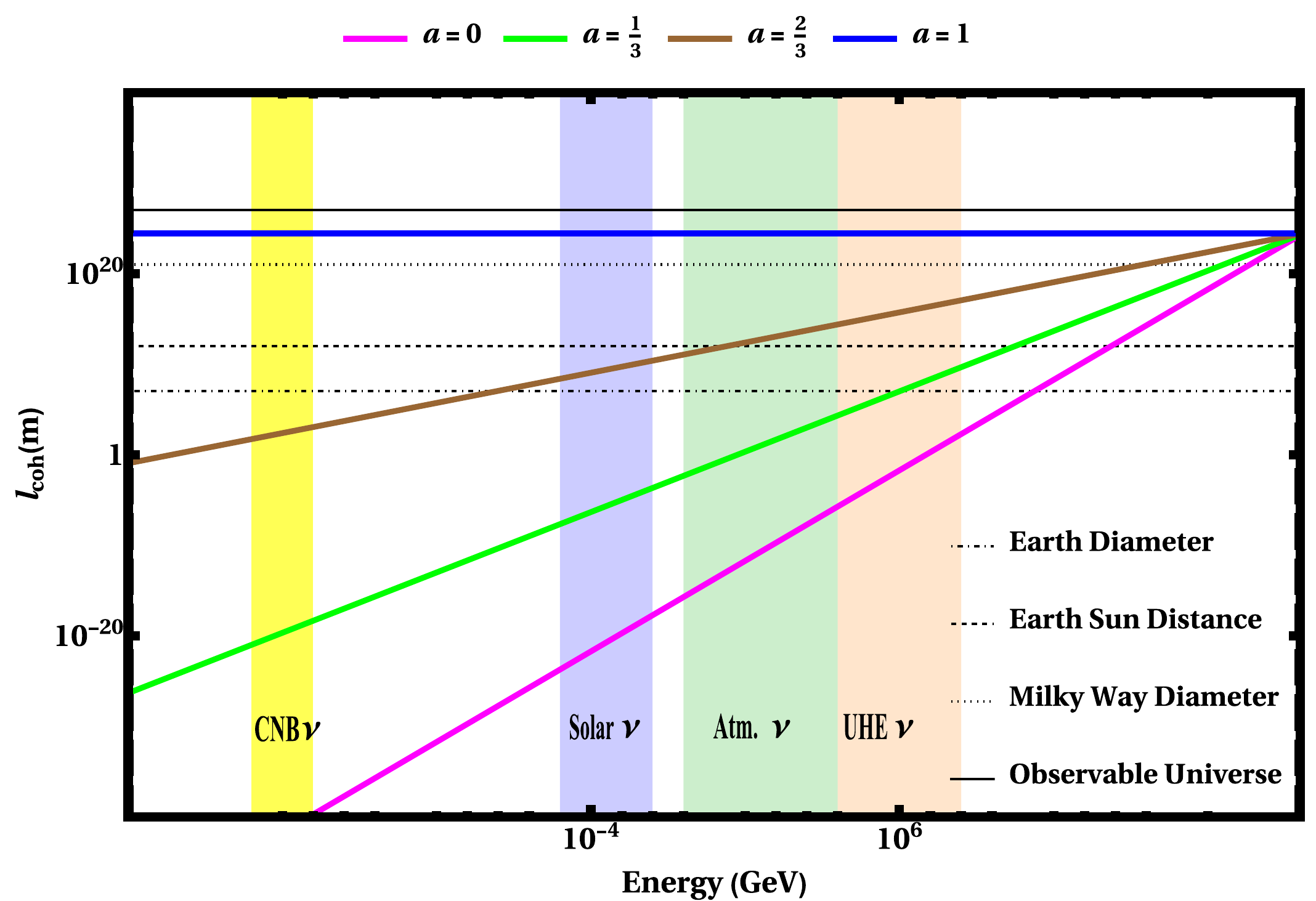}
\caption{The dependence of coherence lengths on energy is shown. The bands represents energy ranges of neutrinos coming from a definite source.}
\label{fig:cohlen}
\end{figure}
In Fig.~\ref{fig:cohlen}, the coherence lengths corresponding to various values of $a$ are illustrated. As expected, these lengths exhibit a steady increase with energy until they reach the Planck energy. For relic neutrinos with energy $\mathcal{O}(10^{-13})$ GeV, the coherence length ($l_{\text{coh}}$) for lower values of $a$ is significantly smaller than both the Earth-Sun distance ($\approx 10^{8}$ km) and the Earth's diameter ($\approx 10^{4}$ km), indicating a loss of coherence. Furthermore, as the value of $a$ approaches 1, $l_{\text{coh}}$ increases substantially, surpassing the milky way diameter ($\approx 10^{18}$ km) while staying below the radius of observable universe ($\approx 10^{24}$ km), thereby providing clear evidence of decoherence in this context as well.
In the case of solar neutrinos with energies $\mathcal{O}(10^{-4})$ GeV, coherence lengths clearly follow a trend akin to that of the cosmic neutrino background (CNB) neutrinos. Decoherence signatures become pronounced as the parameter $a$ changes from $0$ to $1$, even at these lower energy levels. In parallel, for atmospheric neutrinos with energies $\mathcal{O}(10^{3})$ GeV, the impact of decoherence manifests over distances well below the Earth's diameter, remaining observable till the propagation length is comparable to the radius of the observable universe for all possible values of $a$. Furthermore, for ultra high energy (UHE) neutrinos of energy $\mathcal{O}(10^{6})$ GeV, it can be seen that only near to the Earth-Sun distance, the neutrinos lose coherence. As $a\rightarrow1$, the effects become more prominent, making it verifiable in observatories on Earth. 
Here we note that the MSW effects~\cite{Wolfenstein:1977ue, Mikheyev:1985zog, Mikheev:1986wj}, which is relevant for neutrinos moving in matter with varying density, may not be significant here since irrespective of the source, e.g., solar or supernova, the neutrinos are going to cover enormous distances to accumulate quantum gravitational effects proportionately. Therefore, for most part of the trajectory it will resemble vacuum oscillations and thus we are neglecting the matter effects in our study.
	

\subsection{Detection Prospects}
\label{subsec:detecpros}

%
As we can infer from Eq.~\eqref{eq:decohpara}, the decoherence strength parameter depends on energy, the model parameter $a$, and the propagation length $L$. Consequently, for a fixed propagation length, the values of $a$ will affect the sensitivity of 
$\Gamma_{ij}$ with the energy, which in turn will correspond to different types of neutrino oscillation experiments where such signatures can be verified. 

For smaller values of $a$, particularly near zero, the decoherence parameter is suppressed by the neutrino energy, making it sensitive enough for various experiments. However, as $a$ approaches 1, the decoherence parameter becomes suppressed by the Planck energy $ E_{P}$, rendering it less favorable for investigation in this regime. Interestingly, for $a=1/2$, neutrino experiments with energy thresholds of a few MeV can also be sensitive enough to meet the current detectable bounds on the decoherence strength parameter. This is particularly relevant for neutrinos propagating over distances of the order of the radius of the observable universe. In this section, we discuss the prospects for verifiable decoherence signatures across various propagation lengths and categorize the viable experiments accordingly. 
\begin{enumerate}
\item \textbf{Earth Diameter:} For a propagation length of about $10^4$ km, achieving the detectable sensitivity bound for the decoherence strength parameter requires an energy threshold of around $10^3$ GeV. This energy level aligns well with the detection capabilities of experiments such as IceCube~\cite{IceCube:2002eys} and its future generation counterpart, IceCube Gen-2~\cite{IceCube-Gen2:2020qha}.
\item \textbf{Earth-Sun Distance:} To meet the detectable sensitivity bound for the decoherence strength parameter across a propagation length of roughly $10^8$ km, an energy threshold of about $10^5$ GeV is essential. Experiments such as KM3NeT~\cite{KM3NeT:2024jji}, IceCube Gen-2~\cite{IceCube-Gen2:2020qha} and ANITA~\cite{ANITA:2010hzc} possess the necessary detection capabilities to explore these energies.

\item \textbf{Milky way Diameter:} In order to attain the detectable sensitivity bound for the decoherence strength parameter over a propagation length of roughly $10^{18}$ km, an energy threshold of approximately $10^9$ GeV is imperative. Experiments such as GRAND~\cite{GRAND:2018iaj} and RNO-G~\cite{RNO-G:2023kag} are equipped with the sophisticated detection capabilities necessary to investigate these energy levels.
\item \textbf{Radius of Observable Universe:} Similarly, for a baseline of $10^{24}$ km, to meet the detectable sensitivity bound for the decoherence strength parameter requires an energy threshold of around $10^{12}$ GeV. Experiments such as the proposed POEMMA~\cite{Krizmanic:2019hiq}, ARIANNA~\cite{ARIANNA:2014fsk} are projected to be able to meet such energy levels. 
\end{enumerate}

At this point, we note that high-energy neutrinos that align with the above mentioned coherence lengths can indeed be generated in powerful cosmic accelerators and other sources. These include Active Galactic Nuclei (AGN), interactions between cosmic rays and cosmic microwave background (CMB) photons, particularly around the Greisen-Zatsepin-Kuzmin (GZK) cutoff, quenched superradiance of black holes, and the later stages of Hawking evaporation of Primordial Black Holes (PBHs), especially those found in the tail of the power law mass distribution~\cite{Klipfel:2025jql, Banerjee:2024nga}.

\subsection{Oscillation Probabilities}
\label{subsec:oscprobs}
%
In this section, we delve into the profound impact of decoherence on the oscillation probabilities of neutrinos across various flavour eigenstates. As neutrinos propagate, flavour oscillations occurs; however, decoherence transforms these oscillation patterns significantly. Rather than halting the oscillations, it introduces crucial damping effects that alter the oscillatory behaviour of flavour conversion probabilities in noteworthy ways. 
\begin{figure}[t]
    \centering
    \includegraphics[width=0.48\linewidth]{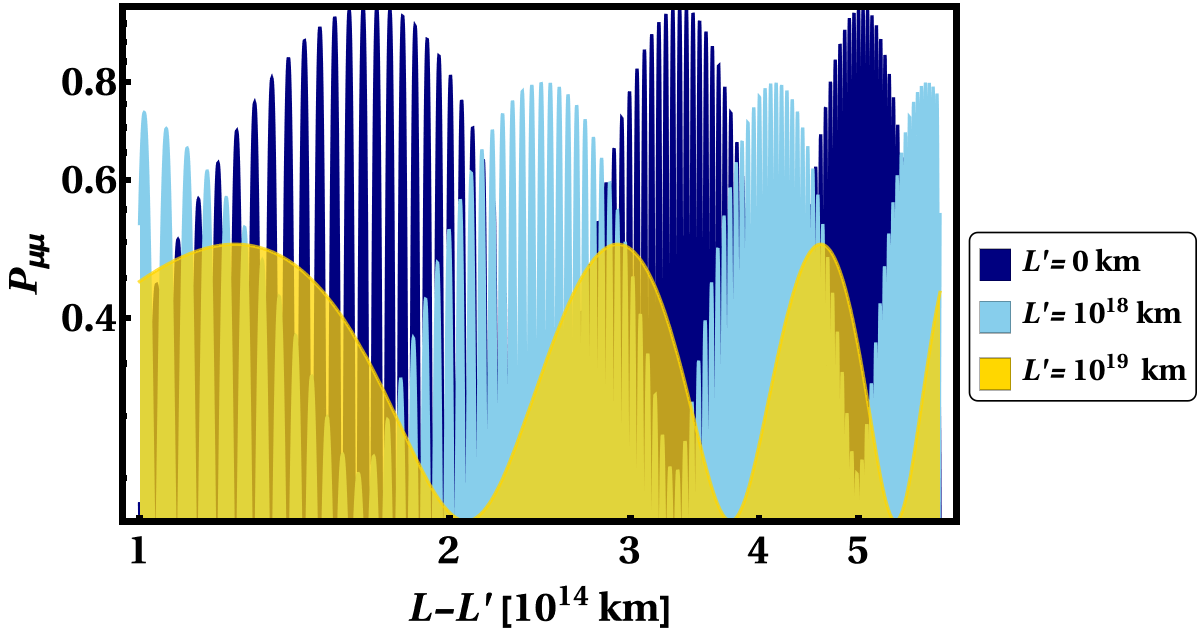}
    \hfill
    \includegraphics[width=0.48\linewidth]{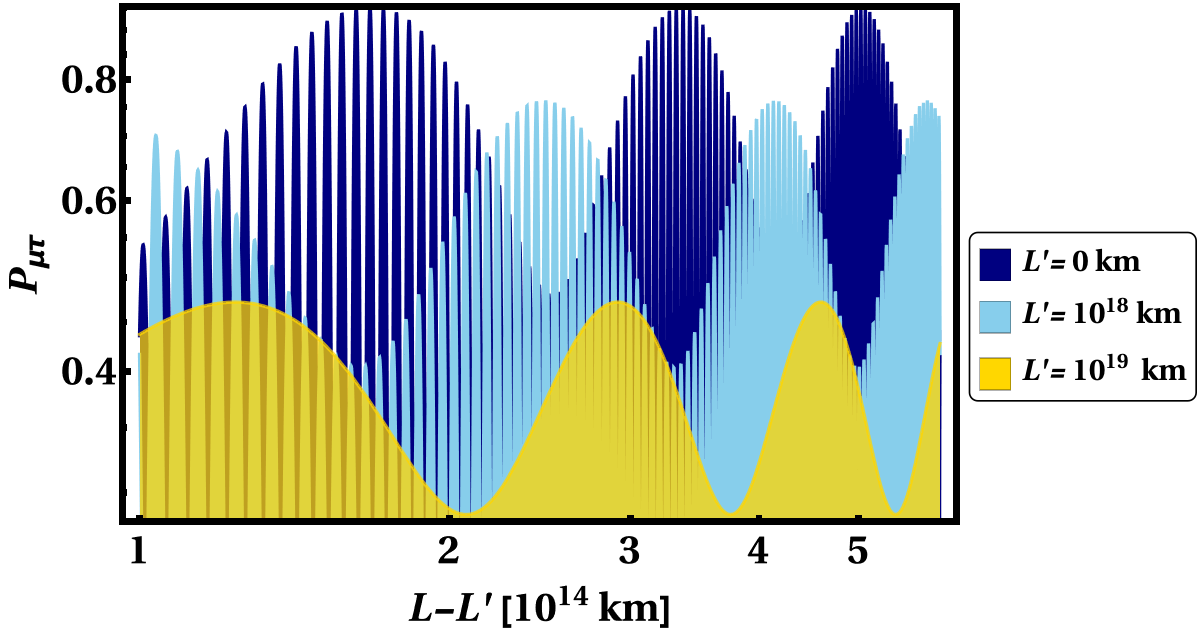}
    
    \vspace{0.4cm}
    
    \includegraphics[width=0.48\linewidth]{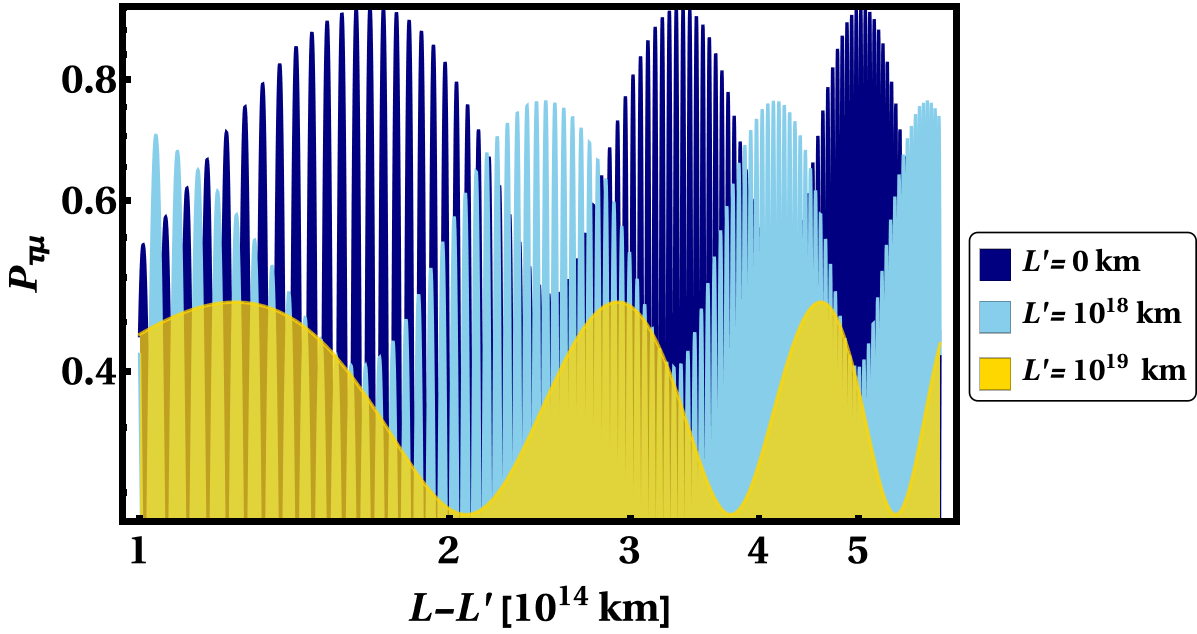}
    \hfill
    \includegraphics[width=0.48\linewidth]{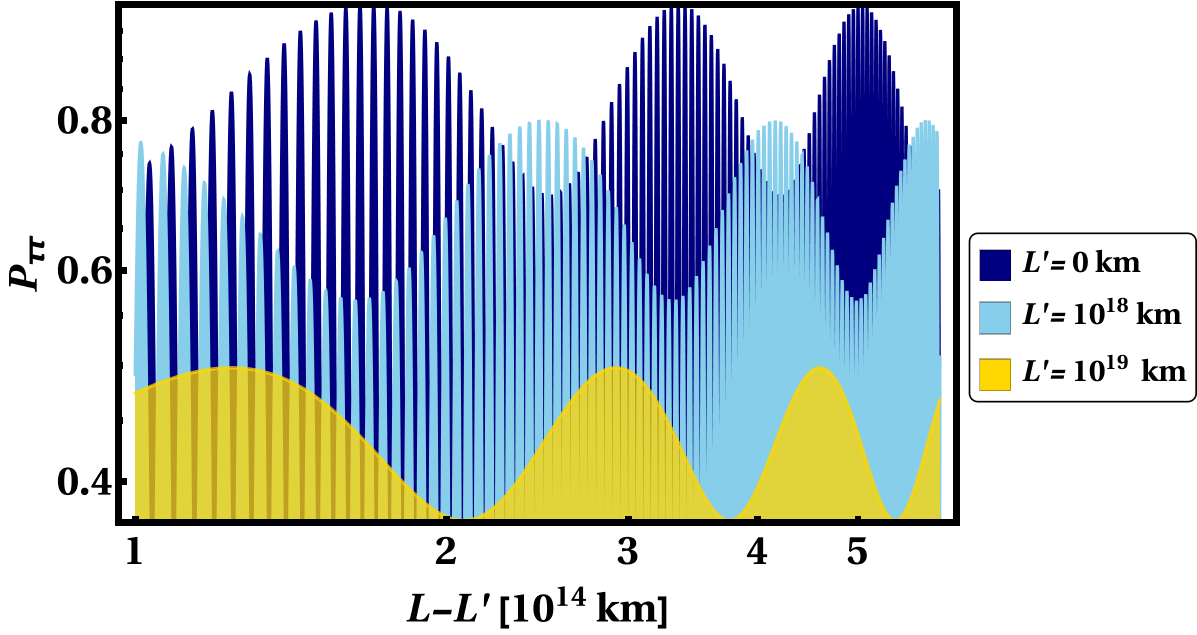}
    
    \caption{Damping signatures on neutrino transition probabilities over time, with $a = 2/3$ and neutrino energy $E_\nu = 100$ PeV.}
    \label{fig:bonuslength}
\end{figure}

In Fig.~\ref{fig:bonuslength}, we show how the decoherence modifies the neutrino oscillation probabilities after different propagation lengths. It is to be noted carefully that we show the oscillation probabilities between the propagation lengths $L^{\prime}$ and $L$, i.e., we start `looking' at the probabilities after the neutrinos have propagated a length $L^{\prime}$ and `look' at it till the neutrinos propagate a distance $L$. This allows us to observe the neutrino oscillation probability in an interval of $L-L^{\prime}$. Furthermore, we consider $L-L^{\prime}\in [10^{14},6\times 10^{14}]\mathrm{~km}$ in order to observe only a few cycles of the neutrino oscillations. We do this to show how the neutrino oscillation probabilities change in a fixed interval after various propagation lengths. It is to be noted here that depicting the oscillation probabilities in the same interval after different propagation length is one of the most efficient way of presenting how the decohrence affects the neutrino oscillation.

In the top left panel of Fig.~\ref{fig:bonuslength}, the muon neutrino survival probability across different propagation lengths is shown.  Notably, as we increase the neutrino propagation length $L'$ given to the neutrino, the damping signatures on the probabilities become strikingly evident. Similarly, on the top right of the panel, the $\nu_{\mu} \rightarrow \nu_{\tau} $ transition probability also highlights the extensive impact of decoherence. The effects for the tau-neutrino family are effectively summarized bottom-left and bottom-right panels of  Fig.~\ref{fig:bonuslength} respectively. Importantly, as the distance travelled by the neutrino increases, we observe the oscillatory (``wiggly") patterns diminishing, highlighting the substantial influence of decoherence in this context.

Furthermore, as neutrinos traverse finite distances, it is evident that the vacuum oscillation probability and the probability that incorporates decoherence begin to diverge. In the context of the $3+0$ flavour formalism, which consists of three active neutrino flavours without sterile components, two distinct mass hierarchies emerge: the Normal Hierarchy (NH), characterized by the condition $m_3^2 \gg m_2^2 > m_1^2$, and the Inverted Hierarchy (IH), defined by $m_2^2 > m_1^2 \gg m_3^2$. The study of decoherence and its effects on neutrino oscillations can shed light on mass hierarchy, which is still an open question, thereby deepening our understanding of the fundamental properties of neutrinos.
\begin{figure}[t]
    \centering
    \includegraphics[width=0.49\linewidth]{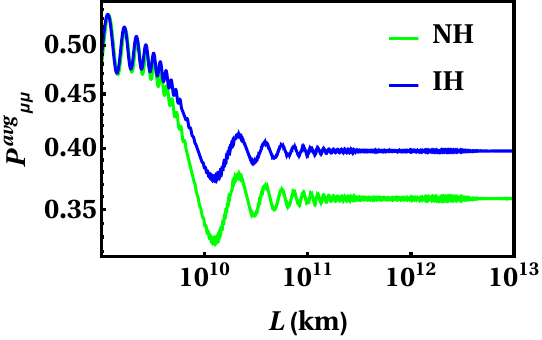}
    \hfill
    \includegraphics[width=0.49\linewidth]{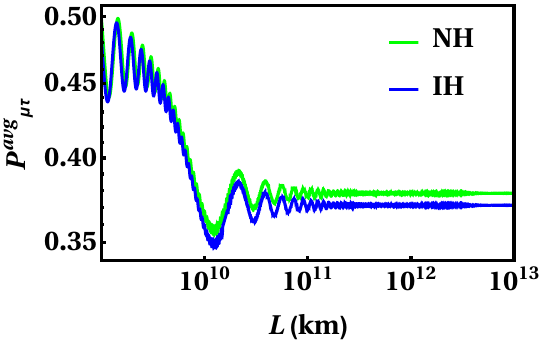}
    
    \vspace{0.4cm}
    
    \includegraphics[width=0.49\linewidth]{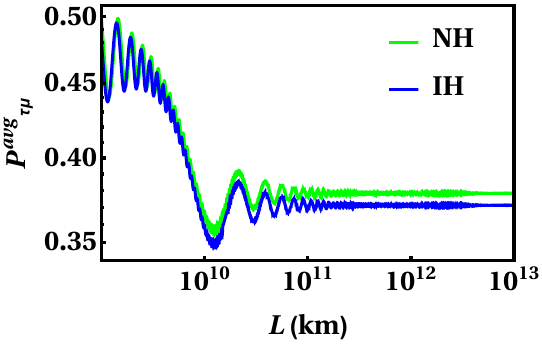}
    \hfill
    \includegraphics[width=0.49\linewidth]{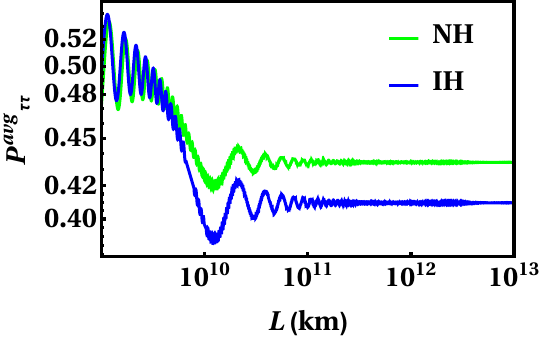}
    
    \caption{The variation of averaged neutrino oscillation probabilities with propagation length is shown with NH and IH taking $a=2/3$.}
    \label{fig:distinctprob}
\end{figure}
In Fig.~\ref{fig:distinctprob}, we show the average oscillation probabilities over the propagation length. To be precise, we show the average oscillation probability integrated over the propagation length $L$ and present them for a few values of $L$. The idea behind this is to take away the oscillatory nature of the probability and focus on the amplitude of the same. Although this method of presentation  still contains some `wigglyness' in the average probability ($P^{\mathrm{avg}}_{\alpha\beta}$), it also enables us to focus on how the decoherence sets in and diminishes the amplitude of transition/survival probabilities.

As illustrated in the top left panel, the effect of decoherence is notably different between the two mass hierarchies NH and IH. After travelling a distance of approximately $10^{10}$ km, the probability points begin to differ for different hierarchies. Similar characteristics for $\nu_{\mu} \rightarrow \nu_{\tau} $ transition probability is advocated by the plot on the top right panel. This significant divergence provides a crucial opportunity to tackle the long-standing neutrino hierarchy problem. The difference of probabilities unequivocally underlines the necessity of identifying and investigating sources at these specific distances. By strategically concentrating our efforts on these distances, the open hierarchy problem can be decisively addressed. Plots on the bottom left and bottom right of Fig.~\ref{fig:distinctprob} reinforce these important characteristics.

\begin{figure}[t]
    \centering
    \includegraphics[width=0.49\linewidth]{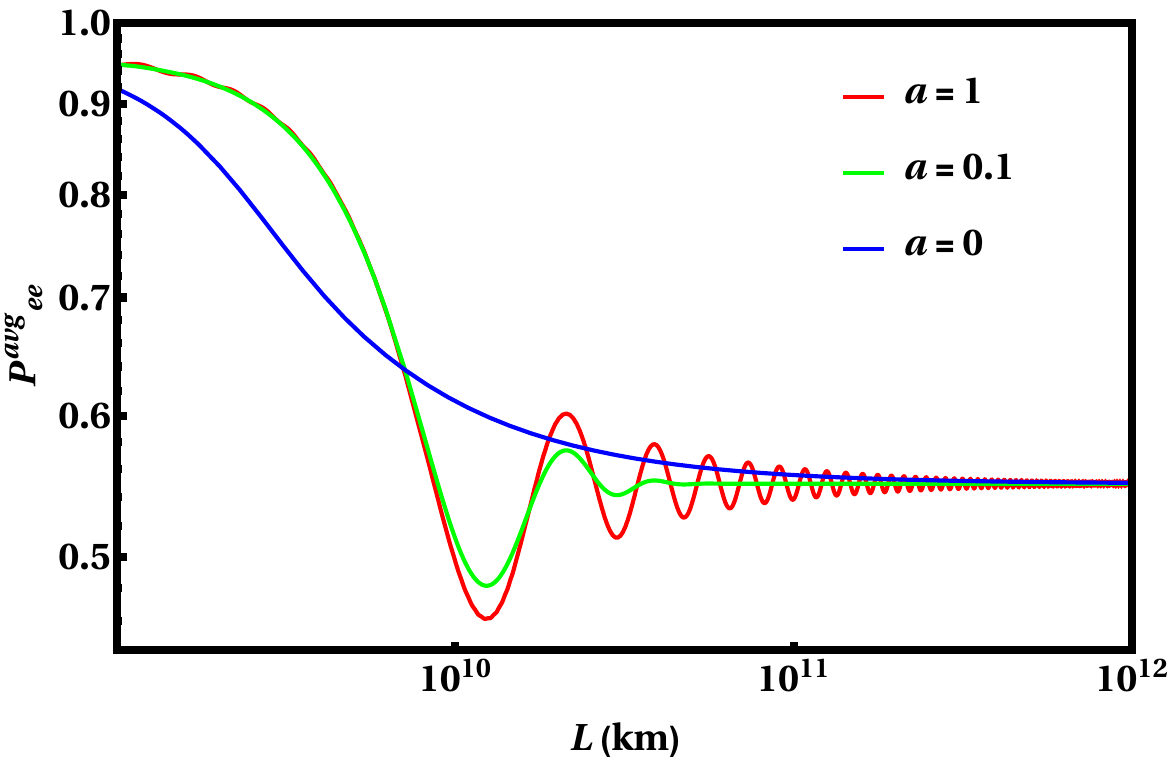}
    
    \vspace{0.5cm}
    
    \includegraphics[width=0.49\linewidth]{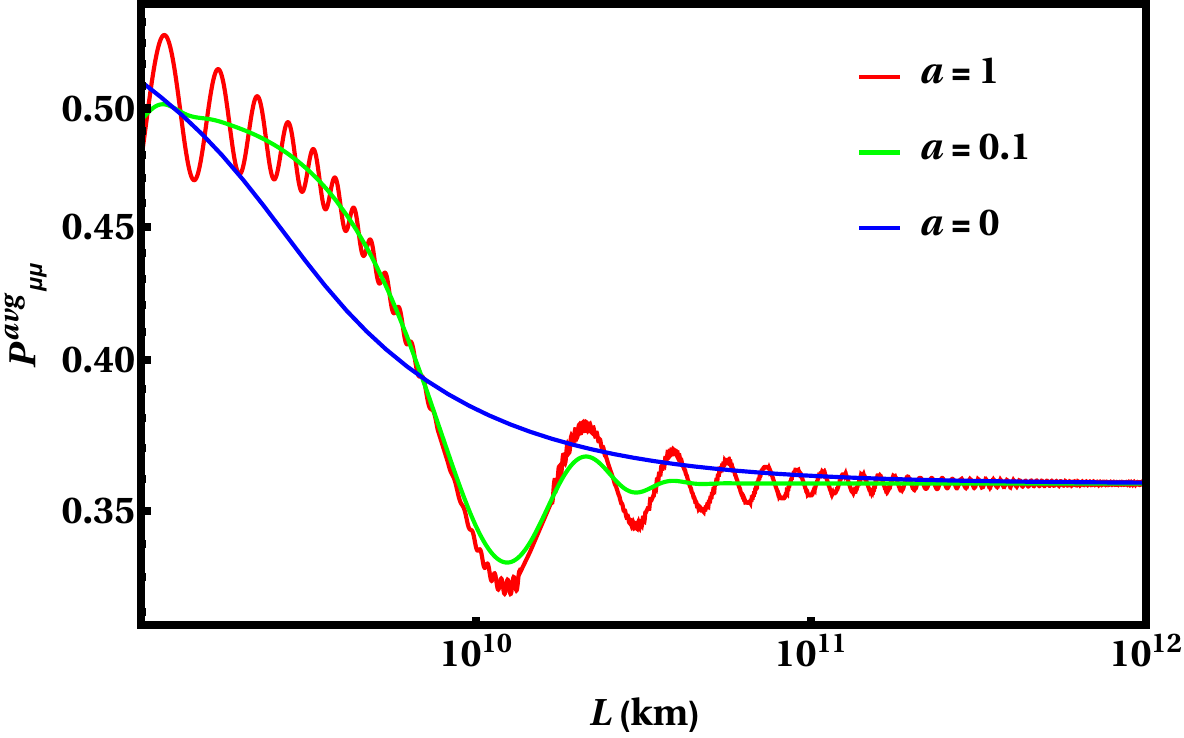}
    \hfill
    \includegraphics[width=0.49\linewidth]{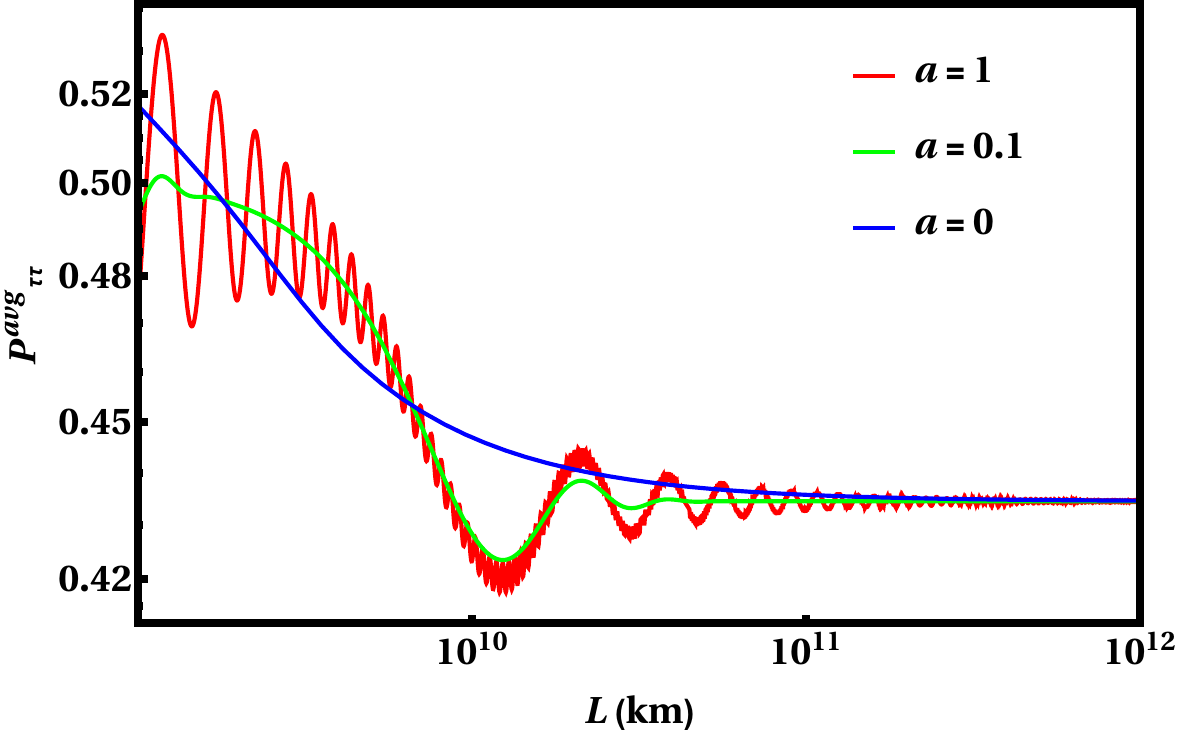}
    \caption{(Top) We show the $\nu_e \rightarrow \nu_e$ survival probability witnessing decoherence. (Bottom-left) Here, $\nu_{\mu} \rightarrow \nu_{\mu}$ survival probability is shown. (Bottom-right) $\nu_{\tau} \rightarrow \nu_{\tau}$ survival probability is shown with $E_\nu=100$ PeV. All the plots are generated with NH.}
    \label{fig:probvariation}
\end{figure}

In investigating neutrino sources that are significantly distant from Earth, it is evident that the oscillatory behaviour of neutrinos diminishes substantially by the time they reach our planet. This characteristic provides an opportunity to examine neutrino oscillation probabilities by averaging over the propagation length. Fig~\ref{fig:probvariation} present the survival oscillation probabilities for the  neutrino across a baseline ranging from $10^{9}$\ km to $10^{15}$\ km. Such an analysis effectively facilitates the selection of neutrinos at selected distances that are particularly pertinent to studies on decoherence.

On the top of Fig.~\ref{fig:probvariation}, the electron neutrino survival probability is plotted against the propagation length, varying the parameter $a$ from $0$ to $1$. In particular, we have chosen $a=0, 0.1, 1$. In the lower propagation length region and for $a \rightarrow 1$, the decoherence effect is almost negligible.  However, as $a \rightarrow 0$, the influence of decoherence becomes more pronounced as the oscillatory behaviour dies out, even at these lower baselines, till the equilibration happens. This highlights the fascinating relationship between decoherence and oscillation probabilities.

On the bottom-left and bottom-right of Fig.~\ref{fig:probvariation}, the muon and tau neutrino survival probabilities further illustrate this trend. For smaller values of $a$, we can observe a notable impact due to decoherence, which adds depth to our analysis. Moreover, within the propagation length range of $10^{9}$ km to $10^{11}$ km, a smooth transition emerges as a direct result of decoherence effects, suggesting a complex yet intriguing dynamic at play. Interestingly, beyond $a=1$ and below $a=0$, the observed behaviour continues to show an insignificant impact of decoherence on all probabilities, providing an avenue for constraining the parameter $a$ and understanding its role. A similar trend can be seen on the bottom-right of Fig.~\ref{fig:probvariation}.
\subsection{Flavour Composition at Neutrino Telescopes}
\label{subsec:flavcomp}
As mentioned before, neutrinos, which propagate through distances before coming to the detectors on Earth, are the best messengers to convey the effect of this decoherence. Furthermore, many of these different sources have been modelled to have different compositions of the neutrinos they produce. Therefore, we focus on the effect of neutrino decoherence on the eventual flux composition of the neutrinos coming from various sources of astrophysical origin~\cite{Anchordoqui:2005is, Beacom:2003nh, Choubey:2009jq}.

We assume these sources will generate fluxes of electron, muon, and tau neutrinos, denoted by  $\Phi^i_e$, $\Phi^i_\mu$, and $\Phi^i_\tau$, respectively. Due to oscillations and, eventually, decoherence, this initial flux composition no longer remains the same as the neutrinos arrive at the detectors on Earth and in space. The measured flux at the detectors is given by the relation,
\begin{equation}
\begin{pmatrix}
\Phi^f_e \\
\Phi^f_\mu \\
\Phi^f_\tau
\end{pmatrix}
=
\begin{pmatrix}
P^{\mathrm{osc}}_{ee} & P^{\mathrm{osc}}_{e\mu} & P^{\mathrm{osc}}_{e\tau} \\
P^{\mathrm{osc}}_{\mu e} & P^{\mathrm{osc}}_{\mu\mu} & P^{\mathrm{osc}}_{\mu\tau} \\
P^{\mathrm{osc}}_{\tau e} & P^{\mathrm{osc}}_{\tau\mu} & P^{\mathrm{osc}}_{\tau\tau}
\end{pmatrix}
\begin{pmatrix}
\Phi^i_e \\
\Phi^i_\mu \\
\Phi^i_\tau
\end{pmatrix}
\label{eq:finflux1}
\end{equation}
where the matrix that is associated with the transformation has neutrino oscillation probabilities as its elements. If we incorporate the effect of coherence loss on neutrino oscillation, then equation  \ref{eq:finflux1} takes the form of:
\begin{equation}
\begin{pmatrix}
\Phi^f_e \\
\Phi^f_\mu \\
\Phi^f_\tau
\end{pmatrix}
=
\begin{pmatrix}
P^{\mathrm{osc+dec}}_{ee} & P^{\mathrm{osc+dec}}_{e\mu} & P^{\mathrm{osc+dec}}_{e\tau} \\
P^{\mathrm{osc+dec}}_{\mu e} & P^{\mathrm{osc+dec}}_{\mu\mu} & P^{\mathrm{osc+dec}}_{\mu\tau} \\
P^{\mathrm{osc+dec}}_{\tau e} & P^{\mathrm{osc+dec}}_{\tau\mu} & P^{\mathrm{osc+dec}}_{\tau\tau}
\end{pmatrix}
\begin{pmatrix}
\Phi^i_e \\
\Phi^i_\mu \\
\Phi^i_\tau
\end{pmatrix}
\label{eq:finflux2}
\end{equation}
Once we have an initial set of flux compositions, we can readily evaluate the final flux compositions at the detectors. Now, we explore all possibilities of initial flux compositions. We can consider the pionic beam dump-like sources, which have an initial flux composition given by~\cite{Beacom:2003nh},
\begin{equation}
\label{eq:pibeam}
(\Phi^i_e : \Phi^i_\mu : \Phi^i_\tau) = (1 : 2 : 0),
\end{equation}
Another composition may arise from a cosmic ray source, and eventually, they decay to secondary hadrons. If the medium in such sources is opaque to muons, then muon-damped cases arises which result in a composition given by~\cite{Choubey:2009jq},
\begin{equation}
\label{eq:mudamped}
(\Phi^i_e : \Phi^i_\mu : \Phi^i_\tau) = (0 : 1 : 0),
\end{equation}
Another flux composition can result from radioactivity that produces neutrons, which eventually decay to produce electron neutrinos. These are called neutron beam sources and have a flux composition given by,
\begin{equation}
\label{eq:neutronbeam}
(\Phi^i_e : \Phi^i_\mu : \Phi^i_\tau) = (1 : 0 : 0),
\end{equation}
Lastly, at very high energies, charm quarks may decay into leptons, which may produce the initial flux composition given by,
\begin{equation}
\label{eq:charmdecay}
(\Phi^i_e : \Phi^i_\mu : \Phi^i_\tau) = (1 : 1 : 0),
\end{equation}
\begin{figure}[H]
    \centering
    \hspace{-1.5cm}
    \includegraphics[width=0.51\textwidth]{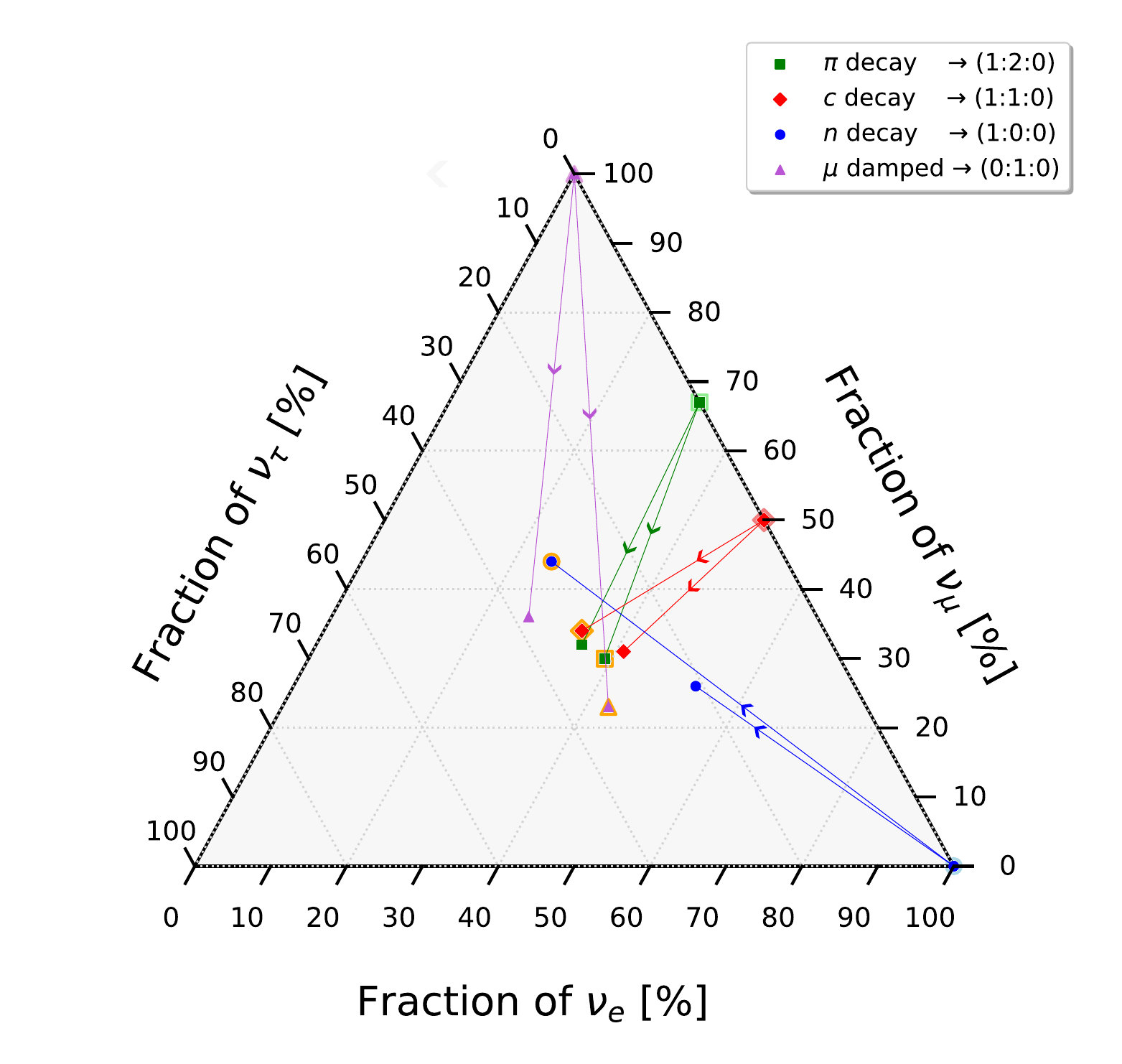}
    \includegraphics[width=0.51\textwidth]{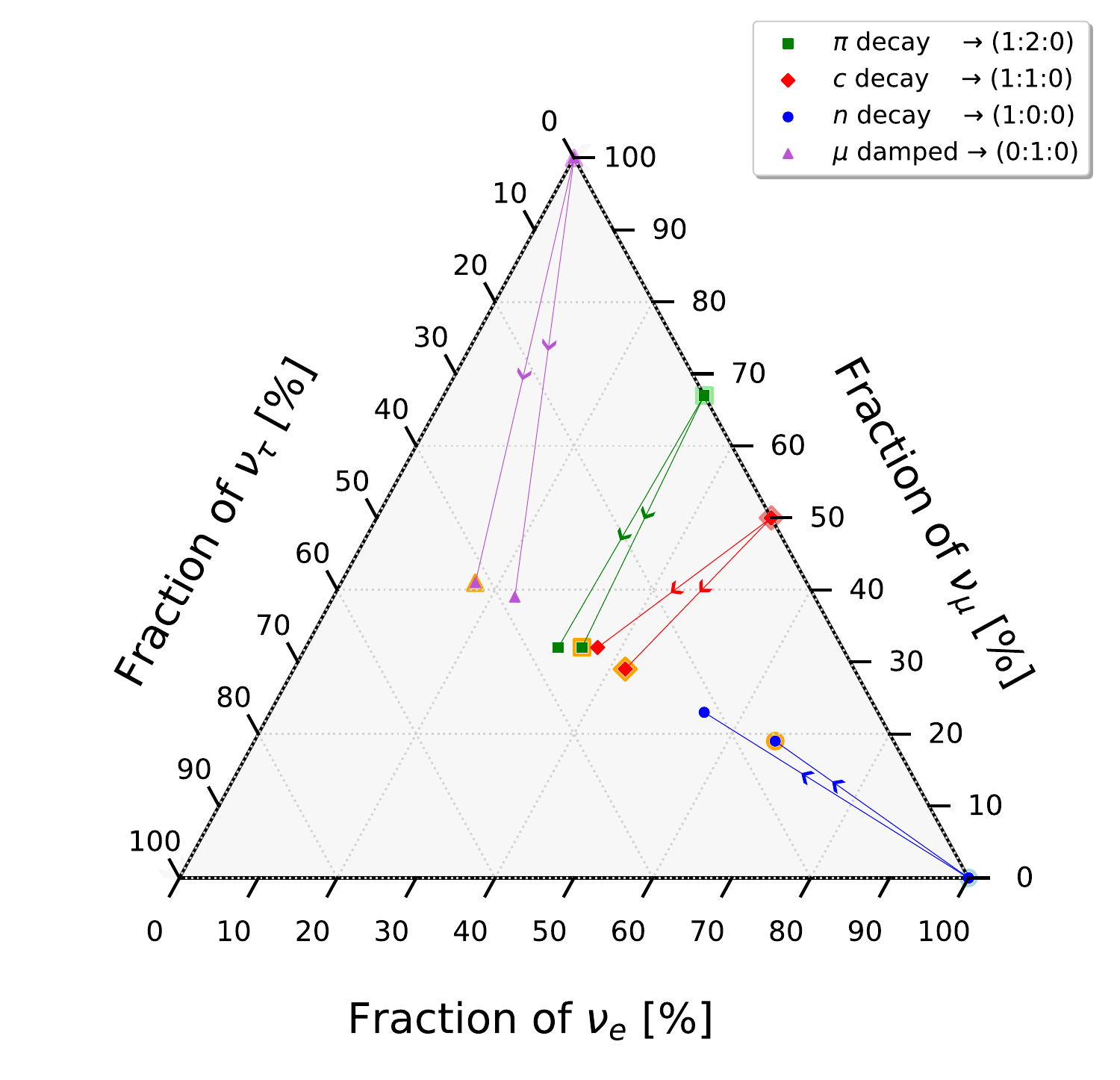}
    \caption{The flavour compositions within the $3+0$ formalism at neutrino telescopes, highlighting the effects of vacuum oscillation and decoherence for a range of initial source flavour compositions. (Left) Flavour compositions at Earth in Normal Hierarchy. (Right) Flavour compositions at Earth in Inverted Hierarchy.}
    \label{fig:probcompare1}
\end{figure}
\begin{table}[H]
\centering
\begin{tabular}{l|ccc|ccc|ccc}
\hline
\multirow{2}{*}{\textbf{Source}} 
& \multicolumn{3}{c|}{\textbf{Initial ($\Phi^i_\alpha$)}} 
& \multicolumn{3}{c|}{\textbf{Final ($\Phi^f_\alpha$) w/o decoherence}}
& \multicolumn{3}{c}{\textbf{Final ($\Phi^f_\alpha$) w/ decoherence}} \\
& $\nu_e$ & $\nu_\mu$ & $\nu_\tau$ 
& $\nu_e$ & $\nu_\mu$ & $\nu_\tau$ 
& $\nu_e$ & $\nu_\mu$ & $\nu_\tau$ \\
\hline
$\pi$ decay     & 1 & 2 & 0 & 0.38 & 0.31 & 0.31 & 0.35 & 0.32 & 0.32 \\
$c$ decay       & 1 & 1 & 0 & 0.34 & 0.34 & 0.32 & 0.41 & 0.31 & 0.28 \\
$n$ decay       & 1 & 0 & 0 & 0.25 & 0.43 & 0.32 & 0.55 & 0.26 & 0.19 \\
$\mu$ damped    & 0 & 1 & 0 & 0.43 & 0.24 & 0.33 & 0.26 & 0.36 & 0.38 \\
\hline
\end{tabular}
\caption{Initial ($\Phi^i_\alpha$) and final ($\Phi^f_\alpha$) neutrino flavour compositions for various production mechanisms, with and without decoherence effects in Normal Hierarchy.}
\label{tab:flavour_compositions1}
\end{table}

\begin{table}[H]
\centering
\begin{tabular}{l|ccc|ccc|ccc}
\hline
\multirow{2}{*}{\textbf{Source}} 
& \multicolumn{3}{c|}{\textbf{Initial ($\Phi^i_\alpha$)}} 
& \multicolumn{3}{c|}{\textbf{Final ($\Phi^f_\alpha$) w/o decoherence}}
& \multicolumn{3}{c}{\textbf{Final ($\Phi^f_\alpha$) w/ decoherence}} \\
& $\nu_e$ & $\nu_\mu$ & $\nu_\tau$ 
& $\nu_e$ & $\nu_\mu$ & $\nu_\tau$ 
& $\nu_e$ & $\nu_\mu$ & $\nu_\tau$ \\
\hline
$\pi$ decay     & 1 & 2 & 0 & 0.34 & 0.33 & 0.33 & 0.32 & 0.35 & 0.32 \\
$c$ decay       & 1 & 1 & 0 & 0.42 & 0.29 & 0.29 & 0.38 & 0.30 & 0.32 \\
$n$ decay       & 1 & 0 & 0 & 0.65 & 0.17 & 0.15 & 0.55 & 0.23 & 0.22 \\
$\mu$ damped    & 0 & 1 & 0 & 0.17 & 0.41 & 0.42 & 0.23 & 0.39 & 0.38 \\
\hline
\end{tabular}
\caption{Initial ($\Phi^i_\alpha$) and final ($\Phi^f_\alpha$) neutrino flavour compositions for various production mechanisms, with and without decoherence effects in Inverted Hierarchy.}
\label{tab:flavour_compositions2}
\end{table}

As seen from the left of Fig.~\ref{fig:probcompare1} and Tab.~\ref{tab:flavour_compositions1}, the flavour compositions of neutrinos change as they travel from their source to Earth. The expected ratios for three different pre-oscillation initial abundances of (1 : 2 : 0), (1 : 1 : 0), (1 : 0 : 0), and (0 : 1 : 0) are represented by green, red, blue, and magenta markers, respectively. At Earth-based neutrino telescopes, the measured post-oscillation abundances, encircled by a yellow marker, are (0.38 : 0.31 : 0.31), (0.34 : 0.34 : 0.32), (0.25 : 0.43 : 0.32), and (0.43 : 0.24 : 0.23). However, when incorporating decoherence arising due to MDR into the analysis alongside post-oscillation data, the measurements shift to (0.35 : 0.32 : 0.32), (0.41 : 0.31 : 0.28), (0.55 : 0.26 : 0.19), and (0.26 : 0.36 : 0.38), respectively. This indicates a significant and verifiable influence of decoherence on the results. All these measured values correspond to normal hierarchy (NH). While on the right of Fig.~\ref{fig:probcompare1}, we conducted similar measurements and measured values are given in Tab.~\ref{tab:flavour_compositions2} with inverted hierarchy (IH). The post-oscillation observations reveal the flavour compositions as follows: (0.34 : 0.33 : 0.33), (0.42 : 0.29 : 0.29), (0.65 : 0.17 : 0.15), and (0.17 : 0.41 : 0.42) for the four outlined sources detailed in Sec .~\ref {subsec:flavcomp}. Notably, when decoherence is incorporated into the analysis, the observed flavour compositions shift to (0.32 : 0.35 : 0.32), (0.38 : 0.30 : 0.32 (0.55 : 0.23 : 0.22), and (0.23 : 0.39 : 0.38). These results underscore significant distinctions in measurements across the two different hierarchies. This signature of decoherence on neutrino oscillation properties decisively opens new avenues to address the still-open question of neutrino preferred hierarchy. At this point, it is important to acknowledge that, in all the flux composition measurements conducted here, we have chosen to set aside the systematic uncertainties involved.

Furthermore, we can draw a firm conclusion about source selection from both left and right of ~\ref{fig:probcompare1}. In the NH case presented on the left of Fig.~\ref{fig:probcompare1}, the post-oscillation measurements, both without and with decoherence, show that the initial pion decay composition of (1 : 2 : 0) does not exhibit any significant shift. Therefore, neutrinos from these sources are not efficient to extract information regarding MDR induced decoherence. Conversely, in the case of neutron decay with an initial composition of (1 : 0 : 0), the post-oscillation measurements reveal a marked distinction between the scenarios with and without decoherence. Therefore, it is evident that neutrino telescopes must focus on those sources where neutrinos are predominantly produced through the neutron decay mechanism. In Parallel, on the right of Fig.~\ref{fig:probcompare1}, with IH, illustrates how the initial composition resulting from charm quark decay reveals a significant shift between observations post-oscillation with and without decoherence. This critical insight allows us to pinpoint the sources where this channel plays a pivotal role in enhancing neutrino production, particularly for decoherence investigations.

\section{Extension to ($3+1$) flavour scenario}
\label{sec:modelextend}
%
In this section, we discuss relevant probes for studying neutrino decoherence, such as oscillation probabilities, and flavour compositions in the 3+1 scenario. Recent analyses of neutrino oscillation data support the standard three active neutrinos, i.e., $3 + 0$ picture. However, some short baseline anomalies favours the existence of a fourth  sterile neutrino~\cite{ALEPH:1989kcj} with an eV-scale mass~\cite{LSND:2001aii, MiniBooNE:2008yuf, Mueller:2011nm, Mention:2011rk, Huber:2011wv, MiniBooNE:2018esg, MiniBooNE:2020pnu}. If we overlook this potential, we may miss crucial interference effects related to additional parameters creeping in, jeopardizing the accuracy of long baseline neutrino oscillation experiments. Hence, it is vital to explore this possibility to enhance the integrity of neutrino decoherence studies~\cite{Hellmann:2021jyz, Domi:2024muy}. In such a case, the mixing matrix can be expressed as,
\begin{equation}
	U^{3+1} = R(\theta_{34}, \delta_{34}) R(\theta_{24}, \delta_{24}) R(\theta_{14}, \delta_{14}) R(\theta_{12}, \delta_{12}) R(\theta_{23}, \delta_{23}) R(\theta_{13},, \delta_{13}), 
	\label{eq:}
\end{equation} 
where $R(\theta_{ij}, \delta_{ij})$ is a rotation in the $ij$-th plane with an associated phase $\delta_{ij}$. In this work, we assume all the CP phases\footnote{Similar study with finite phases have been carried out in ~\cite{Buoninfante:2020iyr} in a three flavour scenario discriminating Dirac and Majorana phases.} $\delta_{ij}=0$ and take the mixing parameter values as mentioned in the Tab.~\ref{tab:sterile} given below. 
\begin{table}[ht]
\centering
\begin{tabular}{lcc}
\hline
\textbf{Parameter} & \textbf{Best-fit values} \\
\hline
$\sin^2 \theta_{12}$ & $0.317$\\
$\sin^2 \theta_{13}$ & $0.022$ \\
$\sin^2 \theta_{14}$ & $0.009$ \\
$\sin^2 \theta_{23}$ & $0.566$ \\
$\sin^2 \theta_{24}$ & $0.007$ \\
$\sin^2 \theta_{34}$ & $0.116$ \\
$\Delta m^2_{41}$ [$~\mathrm{eV}^2$] & $1.69$\\
\hline
\end{tabular}
\caption{$3+1$ neutrino oscillation parameters used for numerical analyses in our study taken from Refs.~\cite{Fiza:2021gvq, Dentler:2018sju}.}
\label{tab:sterile}
\end{table}

As mentioned earlier in Sec.~\ref{subsec:3fdecohform}, here, in this case we can span the needed observables with $4 \times 4$ generalized Gell-Mann matrices and can repeat a similar procedure as outlined in Sec.~\ref{sec:nuOQS} to obtain the MDR-induced master equation in $3+1$ scenario and consequently the relevant observables.

\begin{figure}[t]
	\centering
	\includegraphics[width=0.49\linewidth]{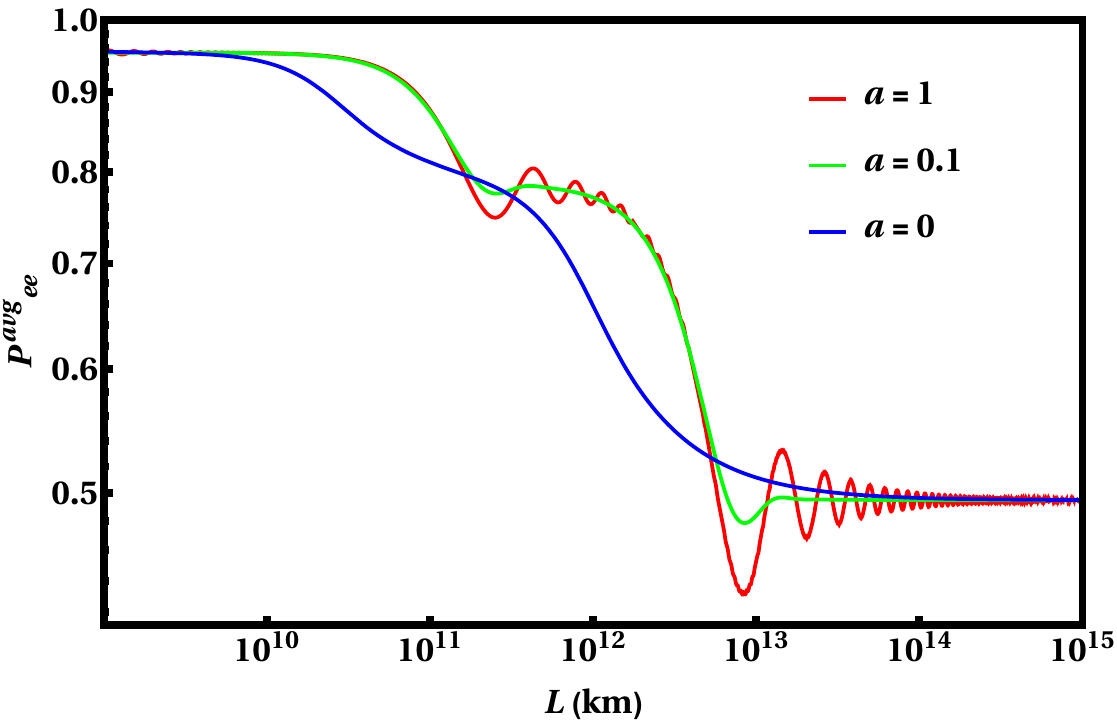}
	
	\vspace{0.5cm}
	
	\includegraphics[width=0.49\linewidth]{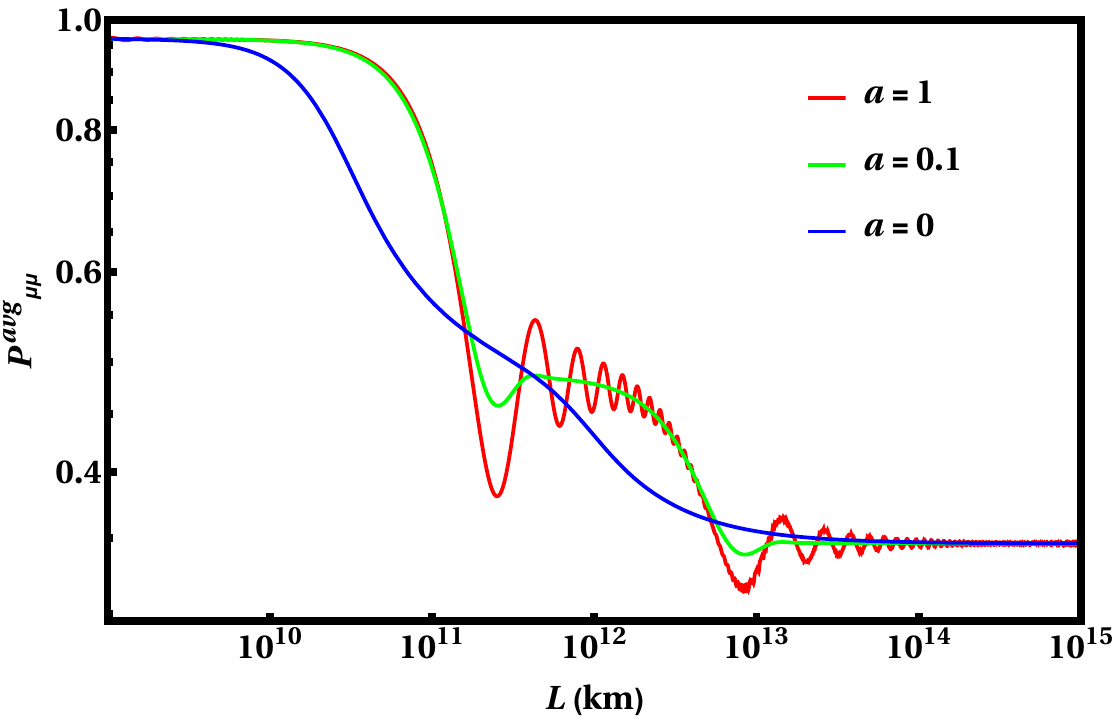}
	\hfill
	\includegraphics[width=0.49\linewidth]{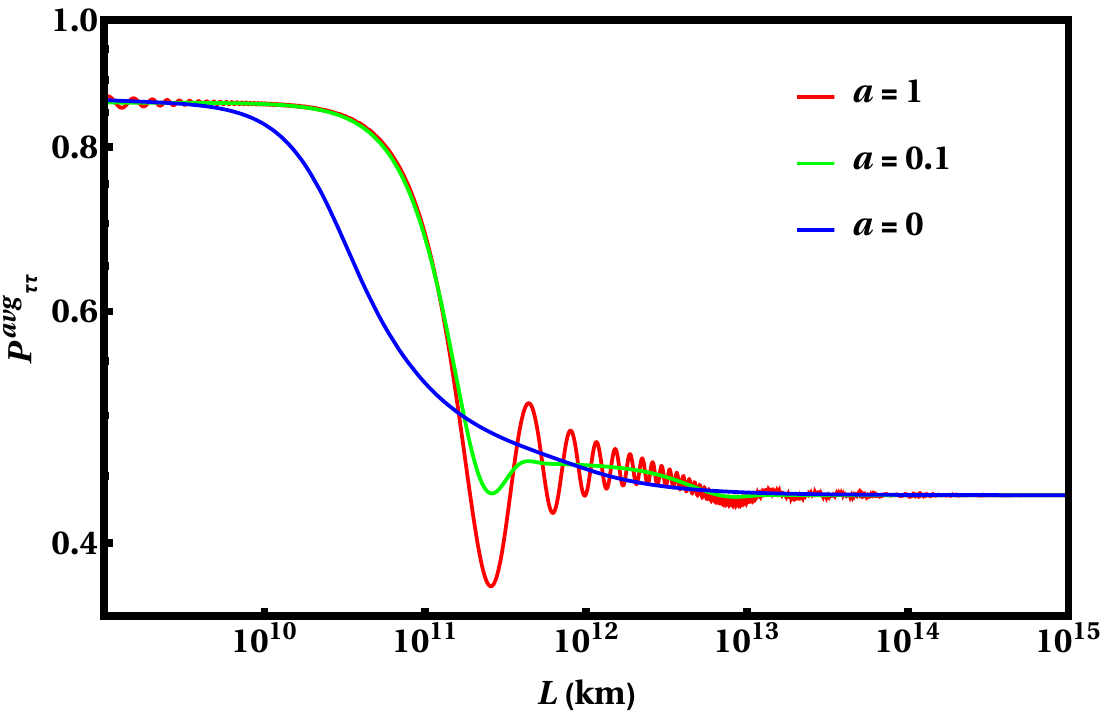}
	\caption{(Top) We show the $\nu_e \rightarrow \nu_e$ survival probability witnessing decoherence varying the propagation length from $1$\ km to $10^{15}$\ km. (Bottom-left) Here, $\nu_{\mu} \rightarrow \nu_{\mu}$ survival probability is shown. (Bottom-right) $\nu_{\tau} \rightarrow \nu_{\tau}$ survival probability is shown. In all the cases, the probabilities are averaged out over baselines with $E_{\nu}=100$ PeV.}
	\label{fig:sterileprobvariation1}
\end{figure}
On top-left panel of Fig.~\ref{fig:sterileprobvariation1}, the averaged electron-neutrino survival probability is shown with the propagation length. It is evident that, at low traversed lengths and for $a \rightarrow 0$, the survival probability is unaffected by decoherence. However, at higher propagation lengths, as  $a  \rightarrow 1$, decoherence effects become prominent and characterized by a significant drop in survival probability followed by equilibration. This drop is distinctly non-smooth, featuring oscillatory characteristics most noticeable at lower propagation lengths, particularly for UHE neutrinos. This behaviour unequivocally indicates the impact of the additional sterile neutrino.
Similarly, on the bottom-left panel of Fig.~\ref{fig:sterileprobvariation1}, we demonstrate the survival probability from  $\nu_{\mu}$ to $\nu_{\mu}$. In stark contrast to the smooth trend observed earlier in the top of Fig.~\ref{fig:sterileprobvariation1}, an anomalous behaviour is apparent here. The probability does not equilibrate smoothly; instead, it decreases in two steps before attaining equilibration as $a$ varies from 0 to 1. This observation highlights the clear distinction between the $3+1$ and $3+0$ formalisms. Since the frequencies of the neutrino oscillation are proportional to the mass square differences, the presence of the heavy sterile neutrinos leads to a frequency which is much larger than the ones already present in $3+0$ scenario. A similar trend can be seen in the $\nu_{\tau}$ to $\nu_{\tau}$ survival probability, as illustrated on the bottom right panel of Fig.~\ref{fig:sterileprobvariation1}.
In recent years, the study of neutrino flavour compositions at Earth-based neutrino observatories has significantly accelerated, providing critical insights into neutrino oscillation properties. We present a robust method to illustrate the four ($3+1$) neutrino flavour compositions observed on Earth over a range of initial compositions. Importantly, in the context of the $3+1$ formalism, the initial source compositions are distinctly categorized into two types: ($\nu_{e}$ : $\nu_{\mu}$ : $\nu_{\tau}$ : $\nu_{s}=0$) and ($\nu_{e}$ : $\nu_{\mu}$ : $\nu_{\tau}$ : $\nu_{s}=x$), determined by the production of sterile neutrinos at the source. 
Both scenarios are well-established within the literature~\cite{Ahlers:2020miq, IceCubeCollaboration:2024nle, Brdar:2016thq, Coloma:2017ppo, Arguelles:2019tum}. In this work, we consider the former case where no $\nu_{s}$ is produced at the source.
\begin{figure}[t]
	\centering
	\includegraphics[width=0.6\linewidth]{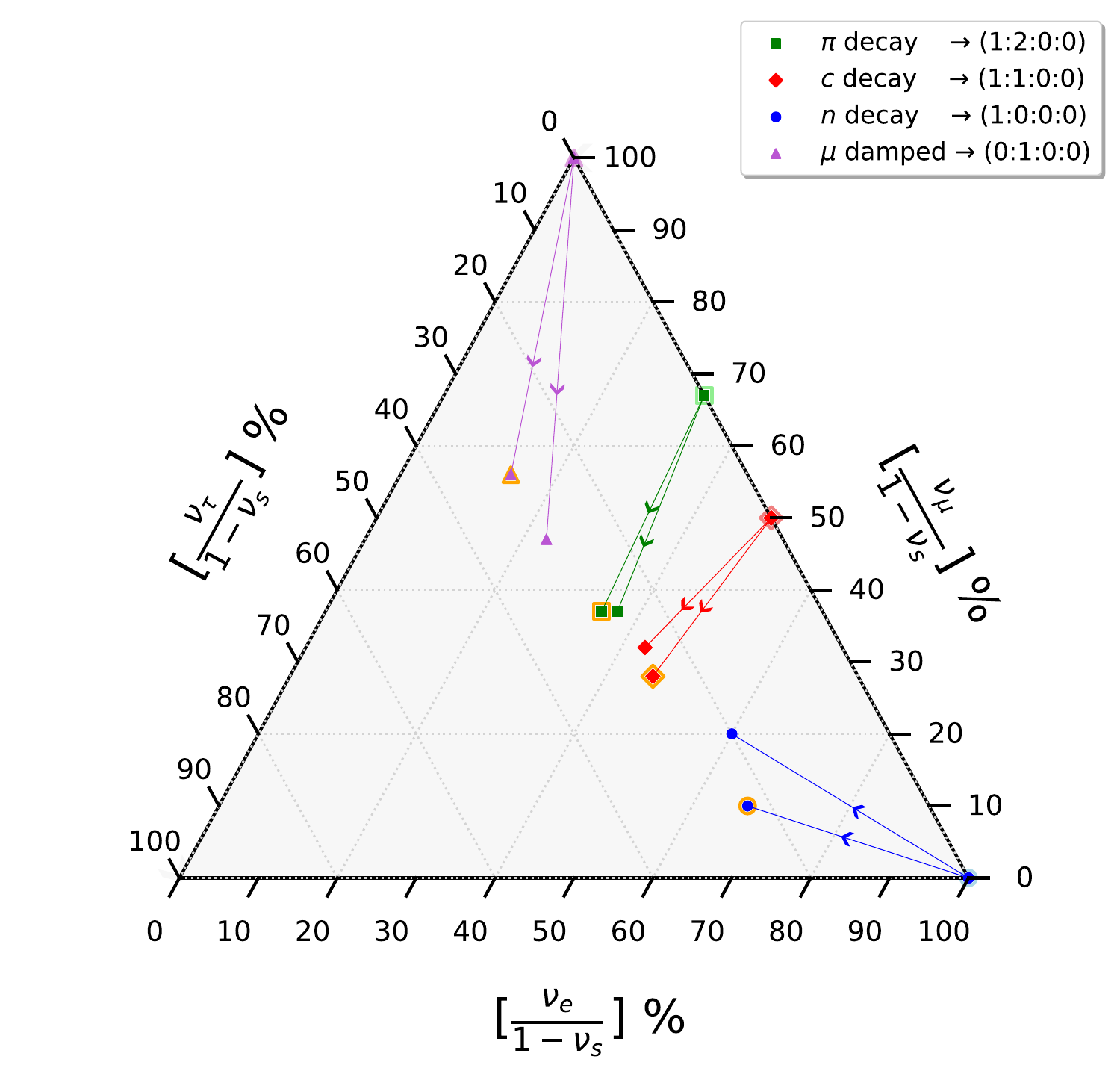} 
	\caption{The flavour compositions within the $3+1$ formalism at neutrino telescopes, highlighting the effects of vacuum oscillation and decoherence for a range of initial source flavour compositions.}
	\label{fig:sterileprobvariation2}
\end{figure}
\begin{table}[H]
	\centering
	\begin{tabular}{l|cccc|cccc|cccc}
		\hline
		\multirow{2}{*}{\textbf{Source}} 
		& \multicolumn{4}{c|}{\textbf{Initial ($\Phi^i_\alpha$)}} 
		& \multicolumn{4}{c|}{\textbf{Final ($\Phi^f_\alpha$) w/o decoherence}} 
		& \multicolumn{4}{c}{\textbf{Final ($\Phi^f_\alpha$) w/ decoherence}} \\
		& $\nu_e$ & $\nu_\mu$ & $\nu_\tau$ & $\nu_s$
		& $\nu_e$ & $\nu_\mu$ & $\nu_\tau$ & $\nu_s$ 
		& $\nu_e$ & $\nu_\mu$ & $\nu_\tau$ & $\nu_s$ \\
		\hline
		$\pi$ decay     & 1 & 2 & 0 & 0 & 0.33 & 0.34 & 0.25 & 0.08 & 0.34 & 0.34 & 0.24 & 0.08   \\
		$c$ decay       & 1 & 1 & 0 & 0 &  0.43 & 0.27 & 0.24 & 0.05 & 0.40 & 0.30 & 0.22 & 0.08   \\
		$n$ decay       & 1 & 0 & 0 & 0 & 0.65 & 0.09 & 0.22 & 0.06 & 0.58 & 0.19 & 0.19 & 0.04  \\
		$\mu$ damped    & 0 & 1 & 0 & 0 & 0.13 & 0.51 & 0.27 & 0.09 & 0.20 & 0.42 & 0.27 & 0.09  \\
		\hline
	\end{tabular}
	\caption{Here, we show initial ($\Phi^i_\alpha$) and final ($\Phi^f_\alpha$) neutrino flavour compositions for various production mechanisms, with and without decoherence effects in $3+1$ formalism varying $a$ from $0$ to $1$ and keeping the neutrino energy $1$ PeV and baseline of $10^{12}$ km.}
	\label{tab:flavour_compositions3}
\end{table}

The flavour compositions of the neutrinos $\nu_{e}$, $\nu_{\mu}$, $\nu_{\tau}$, and $\nu_{s}$ are clearly illustrated in a normalized form in Fig.~\ref{fig:sterileprobvariation2} and Tab. ~\ref{tab:flavour_compositions3}. The post-oscillation observations, which exclude decoherence effects and assume $\nu_{s}=0$ at the source, deliver the compositions: (0.33 : 0.34 : 0.25 : 0.08), (0.43 : 0.27 : 0.24 : 0.05), (0.65 : 0.09 : 0.22 : 0.06), and (0.13 : 0.51 : 0.27 : 0.9) for the four initial compositions outlined in Sec.~\ref{subsec:flavcomp}, respectively.
However, when we incorporate the effects of decoherence, the data showcases a significant shift in these values. The revised compositions are (0.34 : 0.34 : 0.24 : 0.08), (0.40 : 0.30 : 0.22 : 0.08), (0.58 : 0.19 : 0.19 : 0.04), and (0.20 : 0.42 : 0.27 : 0.09), showcasing finite differences in both cases. This evidence decisively indicates that the inclusion of heavy sterile neutrinos has a profound impact on flavour composition measurements on Earth, and advancing decoherence studies will undoubtedly yield critical insights into these scenarios.

In passing we would like to highlight the novelties  of our study as compared to the existing literature e.g., ~\cite{Hellmann:2022cgt} where a rigorous generalised analysis was undertaken.
The most significant difference is our explicit extension of the PMNS matrix to incorporate additional active neutrinos. There it is assumed that the $3+1$ PMNS matrix is taken as \( U = U_{\mathrm{PMNS}} \oplus I_{1 \times 1} \), with the presumption of no mixing between active and sterile fermions. In contrast, we explicitly account for the mixing between these two sectors. Moreover, their omission of a definite form of density matrix makes it difficult to verify essential requirements such as trace preservation and other vital properties necessary for a valid physical density matrix.

\section{Summary and Conclusion}
\label{sec:sumandconc}
%
In this work, we quantitatively discuss the possibility of decoherence in neutrino oscillation due to the possible modifications in the dispersion relation motivated by various quantum gravity models.
After briefly discussing the existing treatments of neutrino oscillations, we show the mathematical formalism of describing neutrinos as an open quantum system, where we discuss the possibility of neutrino decoherence in the three and $3+1$ flavour scenarios. We explain the modified dispersion relation and discuss the relevant parameters that contain the vital information regarding the various quantum gravity theories. In this regard, it is to be noted that in order to relate the modified dispersion relation with existing theories, we use a variable $a$ which can take values between $0$ and $1$, where $a=1$ denotes that the modification to the dispersion relation is Planck energy suppressed, whereas $a=0$ signifies that the modification is entirely dependent on the energy of the system. 
We consider the decoherence parameter (and coherence length), oscillation probability, and the flux composition as the main observables. We have shown that for $a=1$ the decoherence parameter and coherence length are independent of neutrino energy, whereas for $a=0$ the decoherence parameter (coherence length) decreases (increases) with energy. Next we show the evolution of the oscillation probability in this decoherence setup in two different ways. In the first case, we show how the oscillation probabilities change with propagation length and we see that for $a=2/3$ and neutrino energy of $100$ PeV, the oscillation probabilities show a significant change after propagating large distances. In order to understand this behaviour further, we also show the oscillation probabilities averaged over propagation lengths for both normal and inverted hierarchy of neutrino masses. We see that the effect of decoherence manifests differently in normal and inverted hierarchy cases. We then show the evolution of the oscillation probabilities for a $100$ PeV neutrino with in a propagation length ranging from $10^{9}$-$10^{15}$ km and constraining $a$ between $0$ and $1$. 
As expected, the decoherence sets in at very low distances for $a=0$, whereas only for $a=1$ the decoherence is negligible for low energies, but for high energies the decoherence becomes dominant, where the survival (transition) probability sees a drop (rise). Finally, we show the flux composition reaching the Earth and the effect of decoherence in that aspect. We see that the effect of decoherence in this case depends on the initial flux composition, the $a$ value, and the mass hierarchy, i.e., for inverted hierarchy, the deviation of the final flux composition due to decoherence is considerably different than the normal hierarchy.

Apart from the three-flavour case (as discussed above), we also show the effect of decoherence on the oscillation probabilities for a $3+1$ scenario. We show that, however, overall the features of the evolution of survival and transition properties look similar to the three-flavour case, the drop (rise) in the survival (transition) probability has an anomalous drop in the $3+1$ flavour case. This drop can be directly linked to the existence of the extra flavour. On the other hand, due to the presence of the extra flavour, the final flux composition is different in the $3+1$ flavour case than that of the three flavour case. 
Overall, we have shown how the modified dispersion relation can lead to detectable deviations in some regions of energy and propagation length. It is worth mentioning here that we proposed MDR-induced modification of the flux compositions of neutrinos originating from individual sources as an avenue to constrain the parameter $a$. However, we have not performed any statistical analysis for the same due to the lack of enough observational data for individual sources of ultra high energy neutrinos. 

In conclusion, we claim that this study paves the way to establish a link between various quantum gravity theories and neutrinos through the MDR, which can be of extreme importance as various experiments with capabilities of detecting higher energy neutrinos will be in action in the upcoming times.

\acknowledgments
BKA acknowledges the University Grants Commission (UGC), Government of India, for financial support via the UGC-NET Junior Research Fellowship. IKB acknowledges the support by the MHRD, Government of India, under the Prime Minister's Research Fellows (PMRF) Scheme, 2022. UKD acknowledges support from the Anusandhan National Research Foundation (ANRF), Government of India, under Grant Reference No. CRG/2023/003769.

\bibliographystyle{JHEP}
\bibliography{NDMDR.bib}

\end{document}